\documentclass[a4paper,11pt]{article}

\usepackage{jcappub} 

\DeclareGraphicsExtensions{.pdf, .png} 
  
\usepackage[normalem]{ulem}

\usepackage[T1]{fontenc} 
\usepackage{import} 
\usepackage{textcomp}
\usepackage{bigfoot}

\usepackage[normalem]{ulem}

\usepackage[acronym, toc]{glossaries} 
\glsdisablehyper

\newcommand{\clearemptydoublepage}{\newpage{\thispagestyle{empty}\cleardoublepage}}

\newcommand{\fermi}{{\sl Fermi}-LAT\ }
\newcommand{\fermip}{{\sl Fermi}-LAT}
\newcommand{\fiso}{{F_{\rm iso}}}

\newcommand{\dnds}{dN/dS }

\title{Extracting the gamma-ray source-count distribution below the Fermi-LAT detection limit with deep learning}

\author[a,b,1]{A. Amerio,\note{Corresponding author.}}
\author[b,c]{A. Cuoco,}
\author[b,c]{N. Fornengo}

\affiliation[a]{Instituto de F\'isica Corpuscular (IFIC), University of Valencia and CSIC, Calle Catedrático José Beltrán 2, 46980 Paterna, Spain}
\affiliation[b]{Department of Physics, University of Torino, Via P. Giuria 1, 10125 Torino, Italy}
\affiliation[c]{Istituto Nazionale di Fisica Nucleare, Sezione di Torino, Via P. Giuria 1, 10125 Torino, Italy}

\emailAdd{aurelio.amerio@ific.uv.es}
\emailAdd{alessandro.cuoco@unito.it}
\emailAdd{nicolao.fornengo@unito.it}

\abstract{{We reconstruct the extra-galactic gamma-ray source-count distribution, or $\dnds$, of resolved and unresolved sources by adopting machine learning techniques. Specifically, we train a convolutional neural network on synthetic 2-dimensional sky-maps, which are built by varying parameters of underlying source-counts models and incorporate  the  \fermi instrumental response functions. The trained neural network is then applied to the \fermi data, from which we estimate the source count distribution down to flux levels a factor of 50 below the \fermi threshold.
We perform our analysis using 14 years of data collected in the $(1,10)$ GeV energy range. The results we obtain show a source count distribution which, in the resolved regime, is in excellent agreement with the one derived from cataloged sources, and then extends as $\dnds \sim S^{-2}$ in the unresolved regime, down to fluxes of $5 \cdot 10^{-12}$ cm$^{-2}$ s$^{-1}$. The neural network architecture and the devised methodology have the flexibility to enable future analyses to study the energy dependence of the source-count distribution.}}

\begin{document}
\maketitle
\flushbottom

\section{Introduction}

A substantial portion of the gamma-ray radiation that we observe from Earth is produced inside our own galaxy, mostly along the galactic plane. At high Galactic latitudes, however, the gamma-ray emission is mostly of cosmological origin.
This extragalactic background radiation (EGB) is the sum of the emission from all the extragalactic gamma-ray sources \cite{fornasa:2015,Ackermann_2015}. Most of the EGB is originated by various types of astrophysical sources which are seen from our viewpoint as point sources. A relevant observable is therefore their differential source-count distribution $dN/dS$ which counts the number of sources at a given integral source ﬂux $S$ (for gamma-ray energies in a specific interval). For bright sources, which are observationally identified and therefore cataloged, the source-count distribution is directly measured by counting the objects in the catalog,
at least above the threshold for which the catalog detection efficiency is equal to 100\%.  
Below this flux threshold, where the detection efficiency is less than 100\%, dedicated Monte Carlo simulations are required to accurately determine the efficiency and use it as a correction to reconstruct the true underlying $\dnds$ \cite{Fermi-LAT:2010tsy,Fermi-LAT:2015otn,Ajello:2015mfa,Marcotulli:2020fpm}.
For all those sources which are too faint to be resolved, their cumulative distribution of photons in the sky defines an almost isotropic cosmic field, conventionally called the isotropic diffuse gamma-ray background (IGRB) or, more precisely, the unresolved gamma-ray background (UGRB). Even though individual sources below the detector flux-threshold cannot be individually seen, it has been shown \cite{Malyshev_2011, Cuoco-1pdf,Zechlin:2016pme,Lisanti:2016jub}  nevertheless, that it is possible to infer their source-count distribution even in this regime, looking at the collective effects of these unresolved sources. This technique, called pixel-count distribution (or 1-point PDF) and pioneered for gamma-rays in \cite{Malyshev_2011}, has been improved in \cite{Cuoco-1pdf,Zechlin:2016pme,Lisanti:2016jub} by employing a pixel-dependent approach, in order to fully explore all the available information and to incorporate the morphological variation of the gamma-ray emission components. Ref. \cite{Cuoco-1pdf} used the first 6 years of \fermi \cite{Fermi-LAT:2009ihh} data to measure the $dN/dS$ for photons in the energy range (1,10) GeV down to fluxes about one order of magnitude below the \fermi detection threshold. In \cite{Zechlin:2016pme,Lisanti:2016jub} the same technique was used to extend the measurement of the $dN/dS$ to several energy bands between 1 and 171 GeV, thus providing information on the energy dependence of the source-count distribution.
This technique has then been used to study the contribution of individual source classes of emitters, in particular blazars \cite{Korsmeier:2022cwp,Manconi:2019ynl,DiMauro:2017ing,Zechlin:2017uzo}.
The same methodology has also been used to characterize the properties of the unresolved sources in the Galactic Center region in relation to a possible signal from dark matter annihilation \cite{Mishra-Sharma:2016gis,Bariuan:2022phe,Lee:2014mza,Lee:2015fea,Linden:2016rcf,Leane:2019xiy,Chang:2019ars,Buschmann:2020adf,Leane:2020nmi,Leane:2020pfc}.
{  A further methodology called Compound Poisson Generator has also been developed \cite{Collin:2021ufc}, in order to handle biases possibly present in the 1-point PDF method.}

In this paper we update the measurement of the $dN/dS$ below the detection threshold to the increased statistics offered by 14 years of \fermi data. However, differently from Refs. \cite{Cuoco-1pdf,Zechlin:2016pme}, we adopt here a method based on machine learning techniques to obtain the $dN/dS$ below the detection threshold. 
A similar methodology has also been used recently to investigate gamma-ray unresolved sources close to the Galactic center \cite{Mishra-Sharma:2021oxe,List:2021aer,List:2020mzd}.  
We train a convolutional neural network (CNN) on synthetic gamma-ray maps, built from a wide variety of source-count distributions and then apply the trained CNN to the  14-year \fermi map for photon energies in the (1,10) GeV band. We show that the CNN is able to reconstruct the $dN/dS$, thus obtaining an updated result which is fully compatible with the one obtained in \cite{Cuoco-1pdf}. The methodology presented here is also meant to be a proof of principle for the adoption of a CNN to the reconstruction of the source-count distribution of the extragalactic sky, with the future aim of properly investigating additional features of the $dN/dS$, like energy correlations and the presence of additional components which could be traced to dark matter.
{  An advantage of the CNN method is that it
avoids the need to calculate complicated and numerically demanding likelihoods, as in the 1-point PDF method.
Furthermore, we will describe an improved version of the treatment of a CNN on a spherical domain which will further optimize the computational aspect.}

Data selection and \fermi map generation is discussed in Section \ref{sec:dataselection}, while Section \ref{sec:generation} discusses in detail how synthetic maps for the CNN training and validation are constructed. Section \ref{sec:architechture-and-training} discusses the neural network architecture that we use and its implementation, including validation and error estimation. Section \ref{sec:results} discusses the analysis of the \fermi map with the trained CNN and presents the ensuing results for the $dN/dS$. Section \ref{sec:conclusions} gives our conclusions.

\section{Data selection}
\label{sec:dataselection}

\begin{table}[t]
\centering
\begin{tabular}{ | l| l|  }
\hline
Healpix order & 6, 7  \\ 
Weeks & $9-745$  \\ 
Emin & 1 GeV \\
Emax & 10 GeV \\
Instrument Response Functions (IRFs) & \verb|P8R3_SOURCEVETO_V3| \\
EVCLASS & 2048 (Source Veto)  \\ 
EVTYPE & 1 (Front)  \\
ZMAX & 90 \\
\hline
\end{tabular}
\caption{Fermi Tools settings used for the 14-year data set analysis. }
\label{tab:settings}
\end{table}

We process the \fermi data through the Fermi Tools suite \cite{Fermitools} to produce a full-sky map of photon counts for gamma-ray energies in the (1-10) GeV  range. We choose this range as a suitable balance between high statistics (which occurs at lower energies) and good angular resolution (which is better the higher the energy) of the detector, as well as being able to confront our results with those of Ref. \cite{Cuoco-1pdf}.

\begin{figure}[t]
\centering
\includegraphics[scale=0.6]{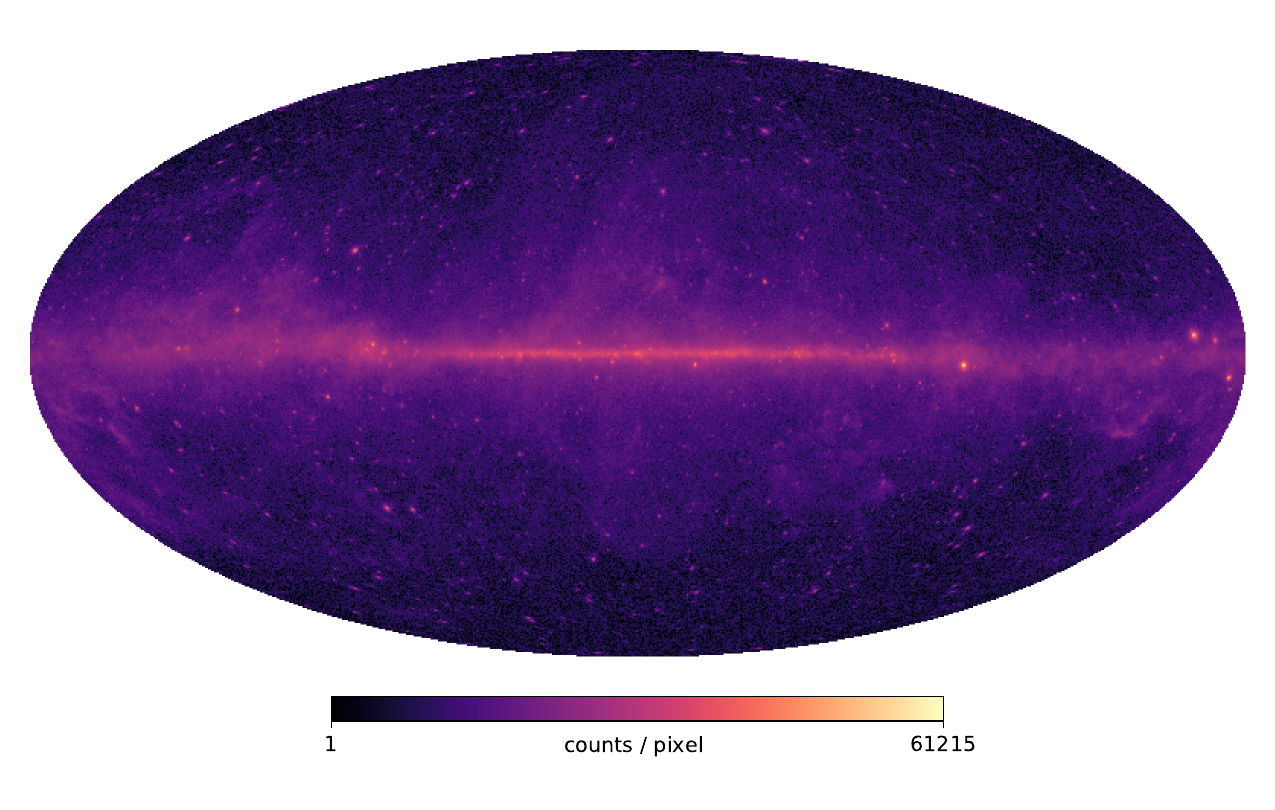}
\caption{\fermi photon-counts map in units of counts per pixel in the 1-10 GeV energy range}  for the 14-years dataset,  at the Healpix resolution $N_{\rm side} = 128$ ($n=7$). 
\label{fig:FermiMap}
\end{figure}

We consider the first 14 years of data collected by the \fermi (specifically, from week 9 to week 745 of operation), and Pass 8 event selection \cite{Fermi-LAT:2013jgq,Bruel:2018lac}. We adopt the \verb|P8R3_SOURCEVETO_V3| instrument response functions (IRFs), event class (EVCLASS) 2048 (\texttt{SOURCEVETO}) and event type (EVTYPE) 1 (\verb|FRONT|). We used standard quality selection criteria, i.e. \verb|DATA QUAL==1| and
\verb|LAT CONFIG==1|.
The atmospheric gamma rays from the Earth limb emission is removed by adopting a cut
on the maximum zenith angle ZMAX of 90 degrees.
\verb|FRONT| events refer to gamma-rays detected by the first layers of the tracking module of the \fermi detector, and are optimised in order to have a better angular resolution, which is a crucial requirement for our analysis. 
The \texttt{SOURCEVETO} class of events achieves good suppression of the charged cosmic-ray (CR) background while still retaining a large event statistics. All those settings are summarised in Table \ref{tab:settings}. Finally, the maps are constructed with equal-area pixels by adopting the Healpix \cite{healpix} pixelisation scheme of order $n=6$ ($N_{\rm side} = 64$ resolution, which corresponds to $0.92^\circ$ pixel side) although for testing we will also use maps with order $n=7$ ($N_{\rm side} = 128$).  
The photon count map, in counts per pixel, for the 14 year data set is shown in Fig. \ref{fig:FermiMap}. This is the map to which we will apply the CNN to extract the source-count distribution of gamma-ray sources. 

\begin{figure}[t]
\centering
\includegraphics[scale=0.6]{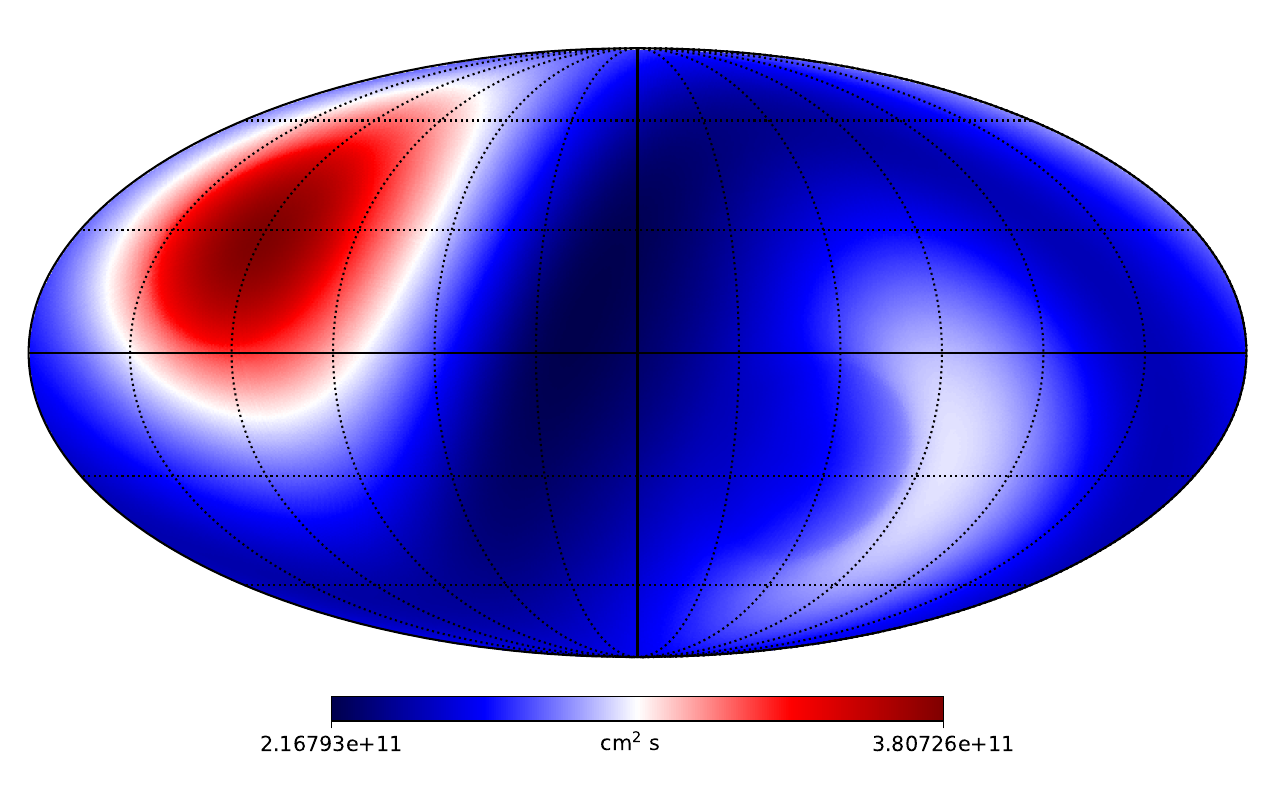}
\caption{\fermi mean exposure map in the 1-10 GeV energy range for the 14-years data set, in units of cm$^2$ s,  at the Healpix resolution $N_{\rm side} = 128$ ($n=7$).} 
\label{fig:exposure}
\end{figure}

\begin{figure}[t]
\centering
\includegraphics[scale=0.6]{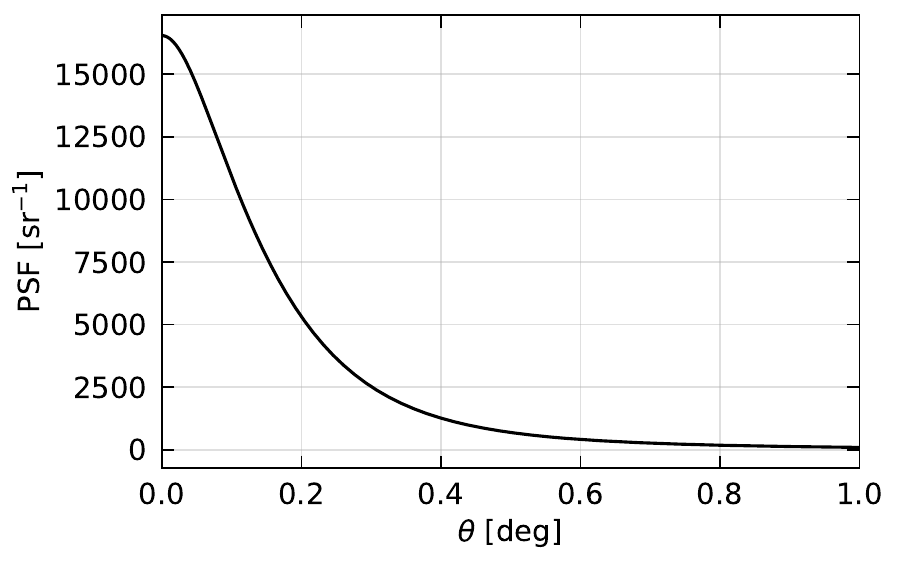}
\caption{Average \fermi point spread function (PSF) for energies in $(1, 10)$ GeV.}
\label{fig:PSF}
\end{figure}

Using the \fermi tools we further extract
the \fermi exposure map and the point-spread-function (PSF) for photon energies in $(1,10)$ GeV, which will be needed to construct the synthetic maps used to train and validate the CNN.
Some care is required, since the PSF is a rapidly varying function of energy.
Following previous works \cite{Cuoco-1pdf,Fornasa:2016ohl},
we build the $(1,10)$ GeV mean PSF by averaging the energy-dependent PSF in the $(1,10)$ GeV energy range with a $E^{-2.4}$ weight, corresponding to the approximate energy spectrum of the high Galactic latitude gamma-ray sky. The average exposure map in the $(1,10)$ GeV energy bin and the averaged PSF profile are shown in Fig. \ref{fig:exposure}  and in Fig. \ref{fig:PSF}, respectively. The \fermi exposure slightly changes with energy, although its variation inside our energy bin is small.

A further dataset we will use is the catalog of resolved sources.
In particular, we will employ the recently released  Data Release 3 of the fourth \fermi catalog of gamma-rays sources (4FGL-DR3) \cite{Fermi-LAT:2022byn}, which increments the previous release to 12 years of data taking and contains 6658 resolved sources. From this catalog, we obtain the (1-10) GeV source-count distribution in the resolved regime, shown in Fig. \ref{fig:dNdS-annotato}. 
The detailed procedure with which we obtain the $\dnds$ from the list of sources is described in Appendix B of \cite{Cuoco-1pdf}.
Here and through the rest of the manuscript $S$ will always refer to an integral flux in the range 1-10 GeV.
The blue points refer to the regime in which the catalog is fully efficient in detecting sources, while the gray points denote sources for which the \fermi sensitivity to point sources starts degrading. 
Our aim is to infer the behaviour of the source count in the unresolved regime, i.e., for source fluxes below $S_{\rm th} \sim 2 \cdot 10^{-10}$ cm$^{-2}$ s$^{-1}$ by adopting a CNN technique.
We stress that in the following the $\dnds$ of resolved sources will be used only for comparison with the result of the CNN analysis, which are completely independent from it.
Thus, the fact that we use 14 years of data while the catalog is based on 12 years does not have an impact on the analysis.

\section{Synthetic map generation}
\label{sec:generation}

In order to instruct the CNN to extract the $\dnds$ from the \fermi photon-count map, we need to produce synthetic analogs of the map, constructed from a wide selection of $\dnds$ and for the same characteristics (detector exposure and PSF, known physical foregrounds) of the data we will be using. These maps will then be partly used to train a CNN and partly used to validate the CNN, to determine its performance and to
estimate the uncertainty on the reconstructed $\dnds$. Finally, we will apply the trained CNN to the \fermi data. 

The gamma-ray maps are modeled as the sum of three contributions:
\begin{itemize}
\item A population of gamma-ray point sources. Since our analysis will be restricted to the high galactic latitude region (in order to minimize in the analysis the impact of the galactic foreground emission, as discussed in the following sections), these sources will mostly be of extra-galactic origin and can therefore be assumed to be distributed homogeneously across the sky. 
The high galactic latitude sky also contains a population of millisecond pulsars which can also be approximately assumed to be isotropically distributed \cite{Fermi-LAT:2013svs}, and thus are ideally included in our phenomenological $\dnds$.
In terms of photon flux, the sources are extracted, for each map realization, from the input $\dnds$ distributions, modeled as in Section \ref{sec:dnds}.

\item Diffuse gamma-ray emission from the Milky Way. Galactic emission is the dominant contribution at low galactic latitudes, with a declining intensity at high latitudes. We model the galactic emission as discussed in Section \ref{sec:galactic}, by adopting a template model normalized to the \fermi data. 
Stability of the results against foreground modeling will be tested through the adoption of different templates.

\item An isotropic background, which integrates all source emissions which are too faint to be processed by the CNN and possible truly diffused emission mechanisms, such as gamma-rays from cosmological cascades from ultra high energy cosmic rays or a possible contribution from dark matter annihilation or decay, as well as a residual instrumental CR background contamination. This term is modeled as a constant.

\end{itemize}
\noindent
A flux map ${\cal M}$ will therefore be obtained as the sum of three contributions:
\begin{equation}
    {\cal M} = {\cal S} + A_{\rm gal} \cdot {\cal G} + \fiso
    \label{eq:map}
\end{equation}
where $\cal S$ is the map due to the underlying distribution of sources (obtained from a $\dnds$ model as described in the next subsection), ${\cal G}$ is a template map for the foreground emission and $\fiso$ is a constant that denotes the isotropic component discussed above. The constant $A_{\rm gal}$ is a normalization constant for the galactic foreground template. 

The count map {$\cal N$} is then obtained by multiplying the flux map by the \fermi exposure map {$\cal E$}, depicted in Fig. \ref{fig:exposure}, and by taking into account the steradian-to-pixel conversion factor. Specifically, the counts in pixel $a$ are obtained as:
\begin{equation}
    {\cal N}_a = {\cal M}_a \cdot {\cal E}_a \cdot \frac{4\pi}{N_{\rm pix}}
    \label{eq:count}
\end{equation}
where the number of pixels in the Healpix scheme is $N_{\rm pix} = 12\cdot N^2_{\rm side}$ with $N_{\rm side} = 2^{n}$ with $n$ the pixeling order.

Since the exposure slightly changes with energy, the above procedure could be improved by performing a convolution in energy of the photon flux with the exposure, instead of directly multiplying the two quantities as performed in Eq. (\ref{eq:count}). However, the gradient in energy of the exposure in the $(1,10)$ GeV energy range is small, and the two procedures would provide very similar results.  We, thus, use the simpler procedure outlined above. 
Eq. \ref{eq:count} also implies that we do not adopt energy dispersion. In the energy range 1-10 GeV the energy dispersion is about 10\% \cite{Fermi-LAT:2013jgq} and may induce normalization changes of similar size.
Thus, as explained in the next section, we will re-derive the galactic foreground normalization with a dedicated fit in order to avoid possible biases in the normalization.

From the count map {$\cal N$}, which contains the number of photon counts in each pixel as determined by the model, we finally produce a Poisson realization {$\cal N_P$}, by extracting in each pixel a photon count from a Poisson distribution with mean equal to the photon number of the corresponding pixel in {$\cal N$}. {$\cal N_P$} constitutes our synthetic map realization, which mimics a count map of an experiment with the \fermi specifications, where the underling model is the one of Eq. \ref{eq:map}.

In the next Sections we discuss in more detail how the components of Eq. \ref{eq:map} are modeled.

\subsection{Differential source-count distribution}
\label{sec:dnds}

\begin{figure}[t]
\centering
\includegraphics[width=0.49\textwidth]{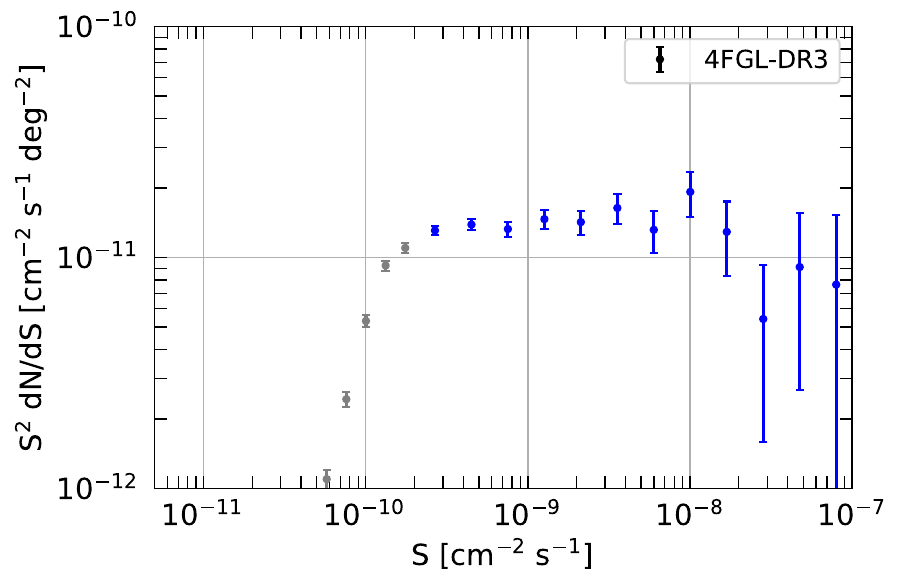}
\includegraphics[width=0.49\textwidth]{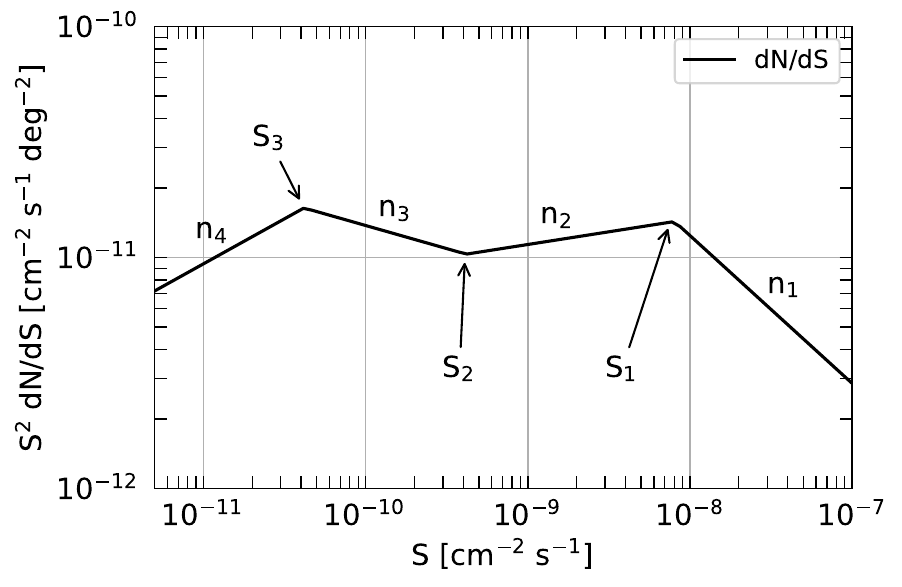}
\caption{Left: The 1-10 GeV $\dnds$ for resolved \fermi sources, obtained from the 4FGL-DR3 \cite{Fermi-LAT:2022byn} catalog. Right: An example of $dN/dS$ with 3 breaks $S_j$ ($j=1,3$) and four slopes, where $n_i$ ($i=1,4$) is the inclination of the slope.}
\label{fig:dNdS-annotato}
\end{figure}

\begin{table}[t]
\centering
\begin{tabular}{lllcc}
\hline
 & Parameter & Prior & Range \\
\hline & $A_{S}$ & log-flat & {$[1,\, 15]$} \\
& $F_{\rm iso}$ & log-flat & {$[0.5 \cdot 10^{-7},\, 7.0 \cdot 10^{-7}]$} \\
\hline &
$S_{1}$ & log-flat & {$[3 \cdot 10^{-9}, 5.0 \cdot 10^{-8}]$ } \\
& $S_{2}$ & log-flat & {$[5 \cdot 10^{-11}, 3.0 \cdot 10^{-9}]$} \\ 
& $S_{3}$ & log-flat & {$[5\cdot 10^{-12},5\cdot 10^{-11}]$} \\
& $n_{1}$ & flat & {$[2,\, 4]$} & \\
& $n_{2}$ & flat & {$[1.80,\, 2.15]$} & \\
& $n_{3}$ & flat & {$[1.5,\, 2.5]$} & \\
& $n_{4}$ & flat & {$[1.0,\, 3.0]$} & \\
\hline 
\end{tabular}
\caption{Ranges of the parameters and their sampling priors adopted in the map generation with $N_b = 3$ breaks. The normalization $A_S$ is in units of $10^7$ cm$^2$ s sr$^{-1}$. $F_{\rm iso}$ is in units of cm$^{-2}$ s$^{-1}$ sr$^{-1}$. The breaks $S_{1}$ $S_{2}$ and $S_{3}$ are given in units of cm$^{-1}$s$^{-1}$. All other quantities are dimensionless.}
\label{tab:prior-dNdS}
\end{table}

Following \cite{Cuoco-1pdf}, the differential source-count distribution $\dnds$ is modeled as a multi-break power-law (MBPL). A MBPL with $N_b$ breaks located at fluxes $S_{j}$ ($j = 1, 2, ..., N_b$) is defined as:

\begin{equation}
\frac{d N}{d S} = A_S \times \left\{\begin{array}{lr}
\left(\frac{S}{S_{0}}\right)^{-n_{1}}, & \quad S>S_{1} \\ \\
\left(\frac{S_{1}}{S_{0}}\right)^{-n_{1}+n_{2}}\left(\frac{S}{S_{0}}\right)^{-n_{2}}, &\quad S_{2}<S \leqslant S_{1} \\ 
\vdots\\\vdots \\
\left(\frac{S_{1}}{S_{0}}\right)^{-n_{1}+n_{2}}\left(\frac{S_{2}}{S_{0}}\right)^{-n_{2}+n_{3}} \cdots\left(\frac{S}{S_{0}}\right)^{-n_{N_{b}+1}}, &\quad S \leqslant S_{N_{b}}
\end{array}\right.
\end{equation}

\begin{figure}[t]
\centering
\includegraphics[scale=0.60]{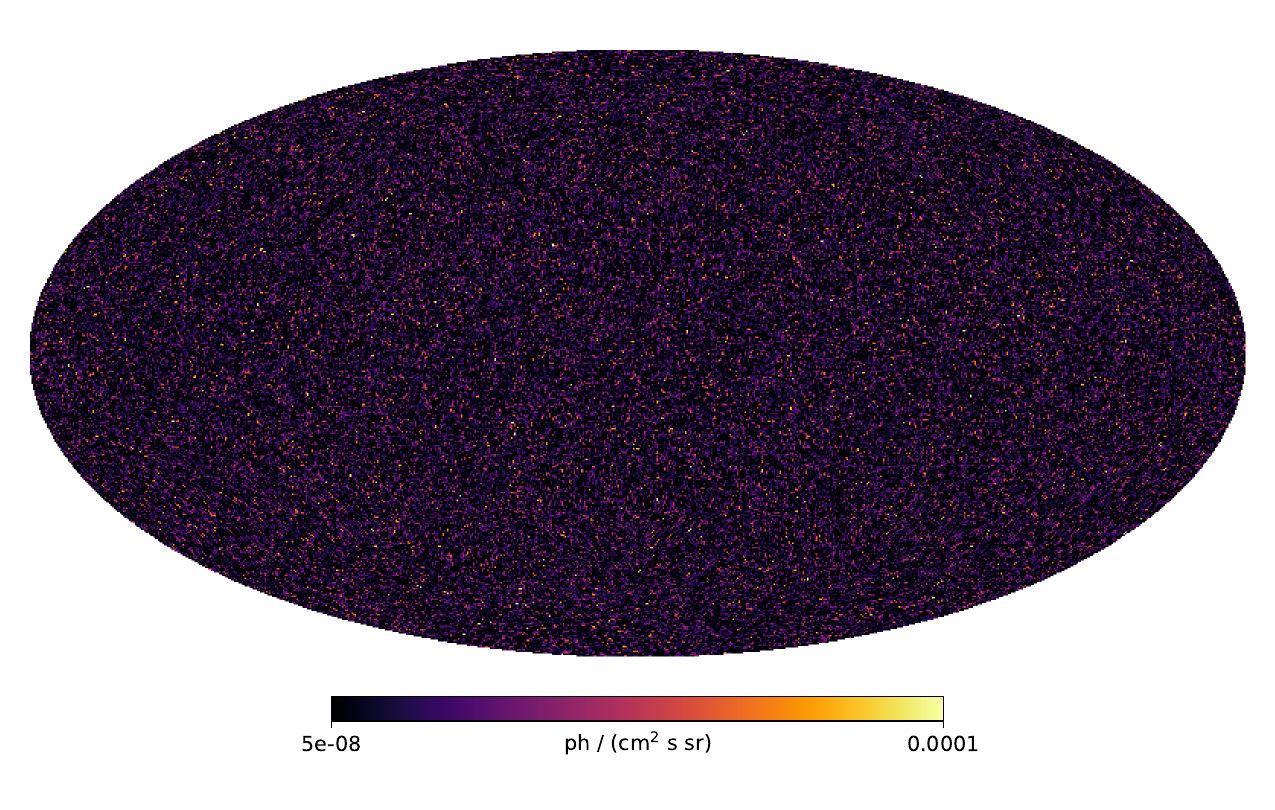}
\includegraphics[scale=0.60]{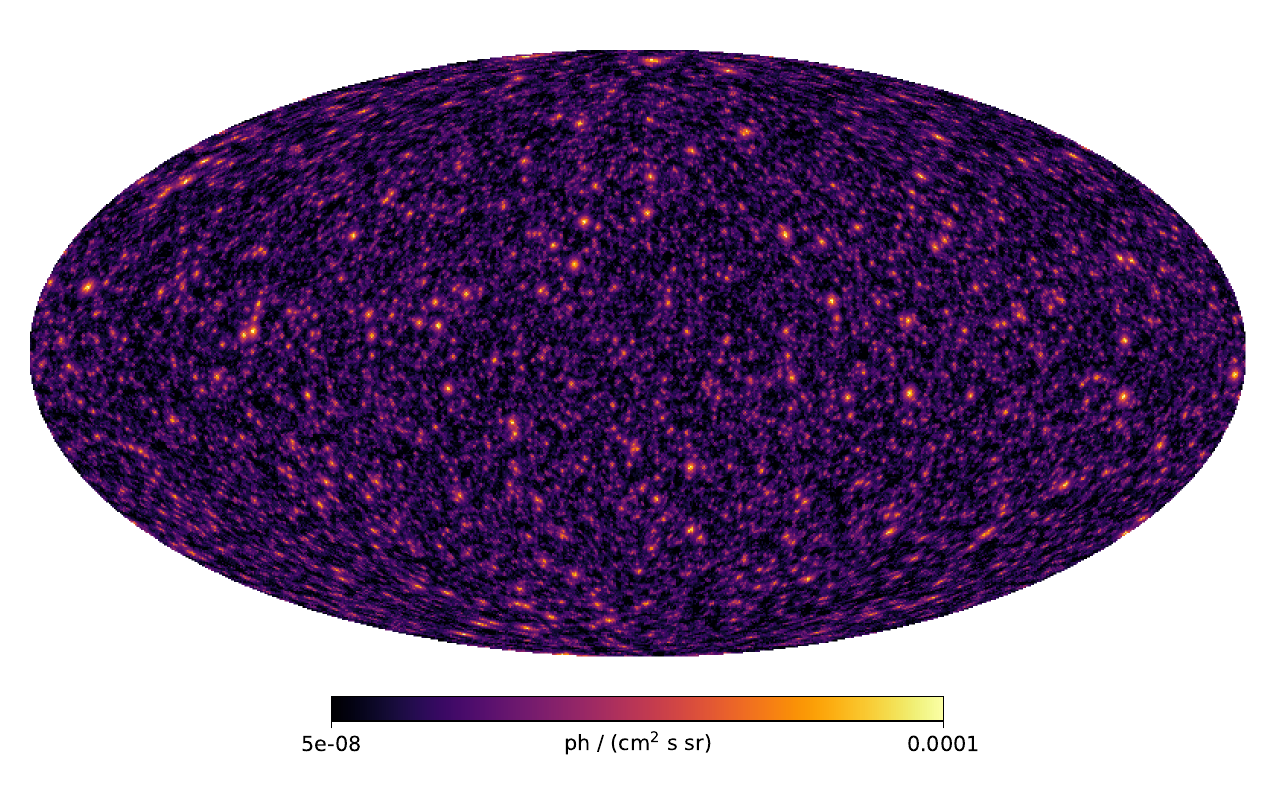}
\caption{Simulated flux map of point sources extracted from a specific source-count distribution, before (upper panel) and after (lower panel) the PSF is applied, in units of photons/(cm$^2$ s sr). The maps are shown at Healpix resolution $N_{\rm side} = 128$ ($n=7$). 
}
\label{fig:point-sources}
\end{figure}

\noindent
where $S_0$ is a reference flux value which we choose to be $S_0 = 3 \cdot 10^{-8} {\rm cm}^{-2} {\rm s}^{-1}$,  and $n_j$ are the power-law indices. $S$ denotes the source flux in the energy interval $(1,10)$ GeV. The $dN/dS$ distribution is normalized with an overall factor $A_S$. 

For definiteness, in this work we adopt 3 breaks and 4 slopes. The ranges in which the parameters are varied to produce different source-count distributions is reported in Table \ref{tab:prior-dNdS}. The ranges have been chosen in order to intercept the expected features of the $\dnds$ and are broad enough to allow for a good variability in the modeling of the source counts, in order to expose the CNN to a wide set of options. In fact, the prior ranges of the three break positions are chosen relative to the \fermi threshold for resolving point sources, $S_{\rm th} \sim 2 \cdot 10^{-10}$ cm$^{-2}$ s$^{-1}$, such that $S_1$, $S_2$, $S_3$ are respectively larger, across and below $S_{\rm th}$. The flux interval over which we will try to reconstruct the $\dnds$ is set as $[5 \cdot 10^{-12}, 1 \cdot 10^{-7}]$ cm$^{-2}$ s$^{-1}$. The upper limit of this interval is set by the brightest sources in the catalog, while the lower limit is set at about 1.6 orders of magnitude below the \fermi threshold for resolving sources. Sources fainter than the lower limit of this interval are assumed to just contribute to the isotropic component.
{  This lower limit, indeed, is also close to the theoretical sensitivity given by the flux of a point source contributing exactly one photon, which we calculate as $F_{\rm sens}= 1/{\cal E}_{av}=3.74 \cdot 10^{-12}$ cm$^{-2}$ s$^{-1}$  where ${\cal E}_{av}$ is the average exposure in the region $|b|>30^\circ$ where we perform the analysis.}

The prior ranges for the power-law slopes are anchored to the results from the 4FGL catalog in the resolved regime, and allow for progressively wider variability moving toward the unresolved regime. Since catalog data exhibit a source count compatible with $S^{-2}$, except for large fluxes, where the slope increases, we adopt priors for $n_1$ and $n_2$ around these behaviours. For $n_3$ and $n_4$, we progressively enlarge the prior ranges, in order to allow for more variability especially in the sub-threshold (unresolved) regime we are interested in. An example of a $\dnds$ generated from this procedure is shown in the right panel of Fig. \ref{fig:dNdS-annotato}, which also illustrates the notation.

The maps {$\cal S$} are generated according to the following procedure:
\begin{enumerate}
\item A $\dnds$ is selected by random sampling of its parameters in the intervals and with the priors defined in Table \ref{tab:prior-dNdS};

\item The number $N_i$ of sources in the $i$th flux bin is calculated as $N_i = 4\pi \, (S_{i, \rm max} - S_{i, \rm min}) \, \dnds|_i$ where $\dnds|_i$ is the value at the center of the $i$th bin. The flux bins are determined by subdividing the interval of interest {$[5 \cdot 10^{-12}, 1 \cdot 10^{-7}]$ } cm$^{-2}$ s$^{-1}$ in $10^3$ log-intervals per decade;

\item A Poisson extraction on $N_i$ determines the number of sources $N_{P,i}$ in each flux interval;

\item The sources are randomly positioned on the sphere by uniformly sampling $\cos\theta$ in $[-1,1]$ and $\varphi$ in $[0, 2\pi]$, $\theta$ and $\varphi$ being the spherical coordinates. The selected angular position is then converted into a pixel position in the Healpix pixeling scheme with the {\tt ang2pix} routine. An example of a point-source map is shown in the upper panel of Fig. \ref{fig:point-sources};

\item Finally, convolution with the \fermi PSF shown in Fig. \ref{fig:PSF} is applied to each point source. Since sources are assumed to be point-like, we simply replace each point source with an extended circular object with a radial profile equal to the PSF, and whose total flux is equal to the one of the original source. Since the image formation process is linear in the intensity, in case of overlapping sources (either because of their spatial position, or because of the PSF smearing) the resulting flux is additive. An example of flux map with the PSF applied is shown in the lower panel of Fig. \ref{fig:point-sources}.
\end{enumerate}

In order to avoid possible biases in the generated maps, due to the resolution of the angular binning, the maps generated with the procedure described above have been produced at a resolution higher than the one adopted for the analysis. We have therefore built our maps with $N_{\rm side} = 128$. These maps have then been downgraded to $N_{\rm side} = 64$ for the analysis.

\subsection{Galactic foreground and isotropic background}
\label{sec:galactic}

\begin{figure}[t]
\centering
\includegraphics[scale=0.6]{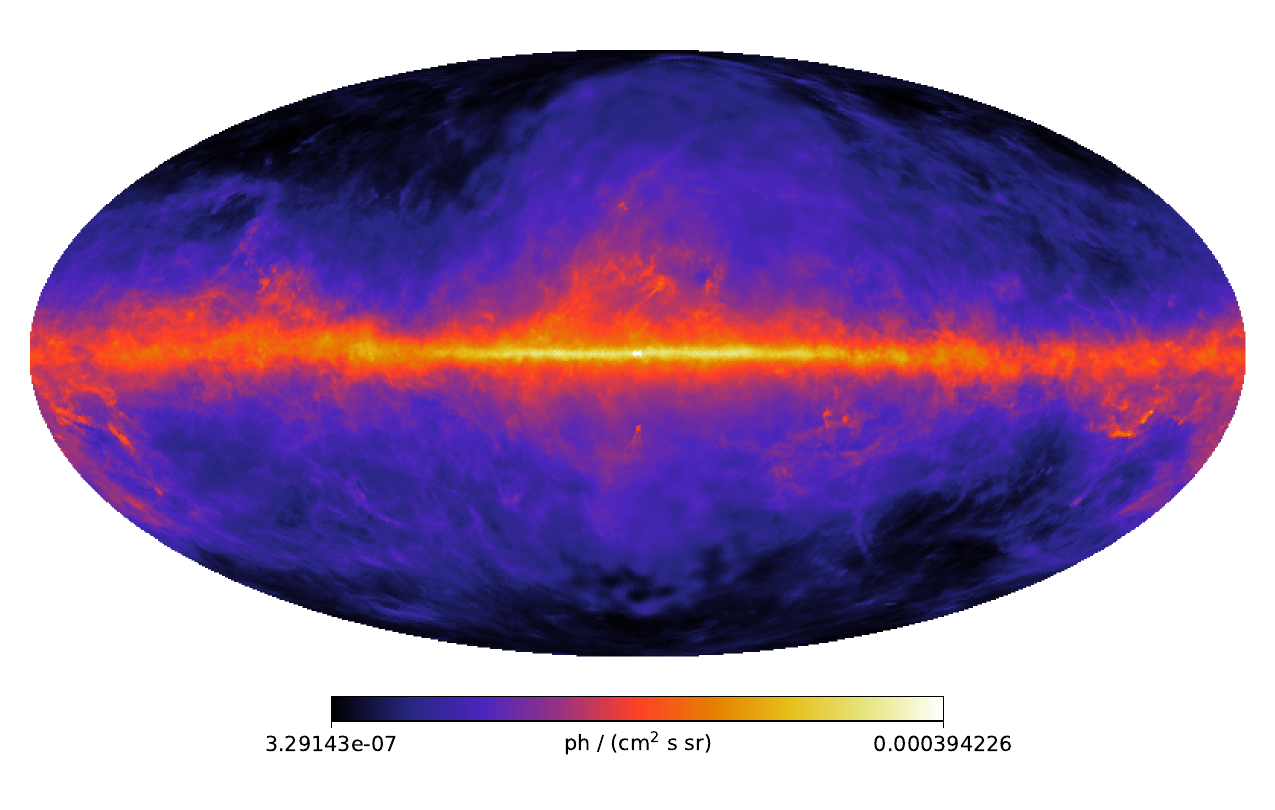}
\caption{$\tt {gll\_iem\_v07}$ galactic foreground model \cite{4FGL}, integrated in the $(1,10)$ GeV energy bin, in units of photons/(cm$^2$ s sr), at the Healpix resolution $N_{\rm side} = 128$ ($n=7$).}
\label{fig:galactic-foreground}
\end{figure}

The galactic foreground is implemented through the adoption of the \fermi template model \verb|gll_iem_v07|\footnote{\verb|https://fermi.gsfc.nasa.gov/ssc/data/access/lat/BackgroundModels.html|}. 
The map of the integrated flux in the $(1, 10)$ GeV energy range is shown in Fig. \ref{fig:galactic-foreground}. To check consistency and stability of our results, we will also adopt a different version of the galactic template, namely model \verb|gll_iem_v05|\footnotemark[\value{footnote}]. The extragalactic isotropic background instead is modeled as a constant flux $\fiso$, which captures all the unresolved emission which is too faint to be associated to the emission modeled through the $\dnds$ (i.e., which refers to fluxes $S<5 \cdot 10^{-12}$ cm$^{-2}$ s$^{-1}$, below our region of interest).

As shown in Fig. \ref{fig:galactic-foreground}, the galactic foreground dominates at low latitudes, although it extends to high latitudes with a smaller flux. For these reasons, low latitudes are unsuitable for the study of the extra-galactic background. Our baseline analysis will therefore apply a mask to the maps to cover the dominant part of the galactic emission, for which we adopt a latitude cut $|b|<30^\circ$. In order to check the stability of the results against the foreground treatment, in Section \ref{sec:results} we will show the results obtained with two additional latitude cuts: $|b|<40^\circ$ and $|b|<50^\circ$.

Given the various approximations we use, instead of using the nominal normalization of the galactic foreground models, we prefer to use a data-driven approach and normalize 
the galactic templates \verb|gll_iem_v07| and \verb|gll_iem_v05| directly to the observed \fermi map. This is realized by performing a maximum-likelihood Poisson fit, using as likelihood:
\begin{equation}
\mathcal{L}=\prod_{a=1}^{N_{\rm pix}} \frac{\lambda_{a}^{k_{a}} e^{-\lambda_{a}}}{k_{a} !}
\end{equation}
where $k_a$ is the number of photons in pixel $a$ of the \fermi count map and $\lambda_a$ the corresponding quantity from the model count-map, given by $\lambda_a = (A_{\rm gal}\ {\cal G}_a + F'_{\rm iso}) \cdot {\cal E}_a \cdot \frac{4\pi}{N_{\rm pix}} $, where $A_{\rm gal}$ and $F'_{\rm iso}$ are the two fit parameters. In performing the fit we mask the sky with the low-latitude cut and with a $2^\circ$ circular mask around each of the sources in the 4FGL catalog. 
We added a prime on the $F'_{\rm iso}$ derived from the maximum-likelihood fit above to explicitly distinguish it from the $F_{\rm iso}$ introduced earlier.
Confronting with Eq. (\ref{eq:map}), the $F'_{\rm iso}$ term denotes the sum of the purely isotropic contribution $F_{\rm iso}$  and the contribution of sources, ${\cal S}$ term, below the 4FGL detection threshold.

\begin{table}[t]
\centering
\begin{tabular}{|l|c|c|c|}
\hline
Foreground template & Latitude cut & $A_{\rm gal}$ & $F'_{\rm iso}$ \\
\hline
\verb|gll_iem_v07| & $|b|<30^\circ$ & $0.888 \pm 0.005$ & $(4.91  \pm 0.04 ) \cdot 10^{-7}$ \\
\verb|gll_iem_v07| & $|b|<40^\circ$ & $0.874 \pm 0.008$ & $(4.90  \pm 0.06 ) \cdot 10^{-7}$ \\
\verb|gll_iem_v07| & $|b|<50^\circ$ & $0.838 \pm 0.013$ & $(5.05  \pm 0.08 ) \cdot 10^{-7}$ \\
\hline
\verb|gll_iem_v05| & $|b|<30^\circ$ & $1.030 \pm 0.006$ & $(2.57  \pm 0.06 ) \cdot 10^{-7}$ \\
\hline 
\end{tabular}
\caption{Values of the $A_{\rm gal}$ and $F'_{\rm iso}$ parameters obtained from a fit to the \fermi map, for different galactic foreground template models and latitude cuts. $F'_{\rm iso}$ is in units of cm$^{-2}$ s$^{-1}$ sr$^{-1}$.}
\label{tab:fore-parameters}
\end{table}

The results of the fit are shown in Table \ref{tab:fore-parameters}, for three latitude cuts for the \verb|gll_iem_v07| template and for $|b|<30^\circ$ for the alternative \verb|gll_iem_v05| model. We see that both the normalisation parameter $A_{\rm gal}$ and $F'_{\rm iso}$ are consistent among themselves for a fixed galactic foreground model, implying consistency with respect to the latitude cut. The model \verb|gll_iem_v05| instead requires a larger normalization, with an ensuing reduced value for $F'_{\rm iso}$ by a factor of 2.
This reflects a certain amount of degeneracy which is present at high galactic latitude between the foreground emission and the isotropic emission, allowed by the uncertainties in the foreground model.
The normalization of the \verb|v07| model different from 1 is related to the fact that we do not include energy dispersion. The \verb|v05| model, instead, has been created originally without including the effect of energy dispersion\footnote{\verb|https://fermi.gsfc.nasa.gov/ssc/data/access/lat/BackgroundModels.html|}, and indeed we obtain a normalization compatible with 1.

\begin{figure}[t]
\centering
\includegraphics[scale=0.6]{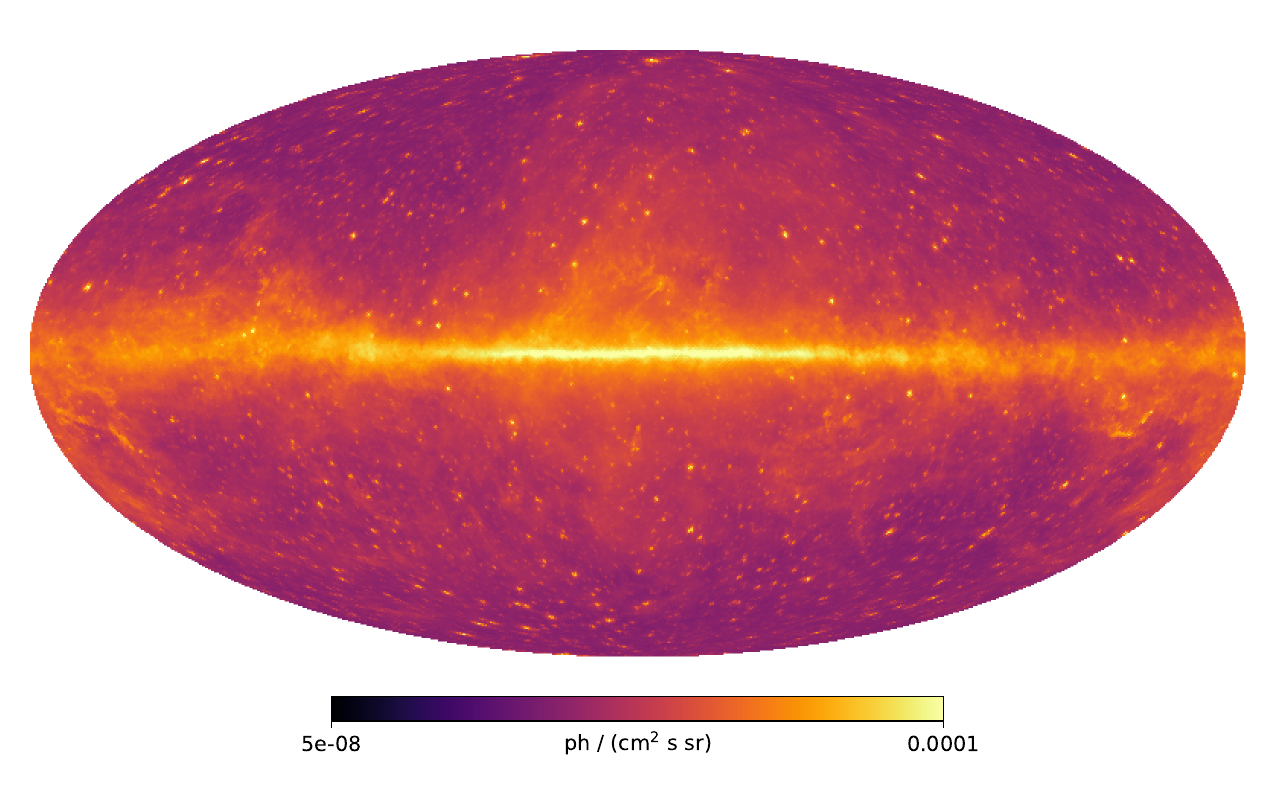}
\caption{The same as in Fig. \ref{fig:point-sources}, with the galactic foreground model normalized with $A_{\rm gal} = 0.886$ and an isotropic background component $F_{\rm iso} = 4 \cdot 10^{-7}$ cm$^{-2}$ s$^{-1}$ sr$^{-1}$. Units are photons/(cm$^2$ s sr).}
\label{fig:map_with_gf}
\end{figure}

Throughout the rest of the analysis we will fix the normalization parameter $A_{\rm gal}$ to the value determined from the above fit. The values of $F'_{\rm iso}$ are, instead, used to define the range of the prior for $F_{\rm iso}$ in Table \ref{tab:prior-dNdS} and {\sl a posteriori} to check consistency of the results.
One illustration of a flux map obtained as the sum of all components is shown in Fig. \ref{fig:map_with_gf}.
Instead of fixing $A_{\rm gal}$, in principle, we could also leave it as a free parameter and instruct the network to predict its value. We found, however, that the above procedure constrains $A_{\rm gal}$ very precisely and it is thus preferable. Furthermore, we also found that fixing $A_{\rm gal}$ provides a better stability of the neural network.  

{  Nonetheless, to better asses the impact of fixing the $A_{\rm gal}$ value, we also performed a further test using different values in the range $(0.78, 1.0)$ (i.e., a range centered on our fiducial value $0.888 \pm 0.005$, thus encompassing many standard deviations away from its central value.)
We found that the $\dnds$ reconstructed by the CNN is largely unaffected by the employed value of $A_{\rm gal}$, while 
the specific $A_{\rm gal}$ choice mostly affect $F_{\rm iso}$,
as expected because of the  high galactic latitude degeneracy between the foreground emission and the isotropic emission. 
Example plots are shown in the Appendix \ref{sec:agal-var}.}

\subsection{Count maps}

\begin{figure}[h]
\centering
\includegraphics[scale=0.6]{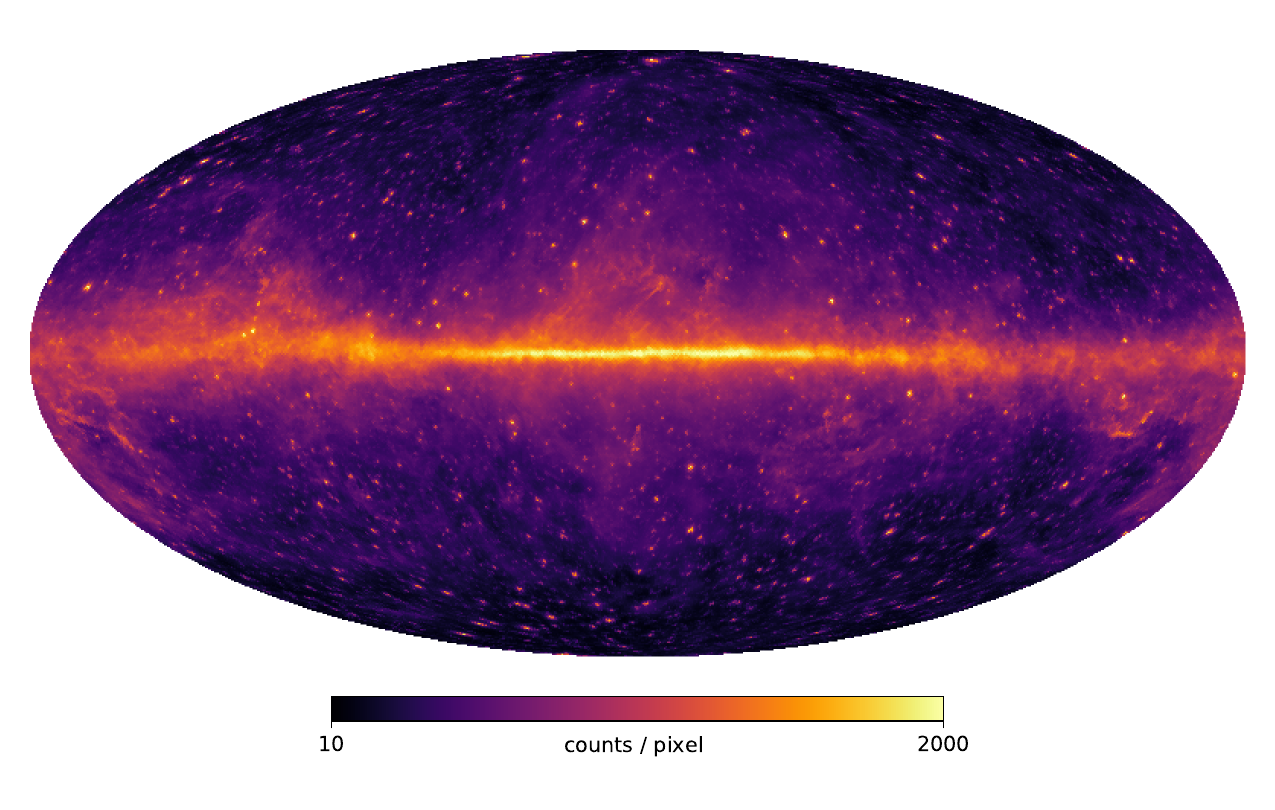}
\caption{Simulated count map in units of counts per pixel,  at the Healpix resolution $N_{\rm side} = 128$ ($n=7$). }
\label{fig:map-total-counts}
\end{figure}

\begin{figure}[h]
\centering
\includegraphics[scale=0.6]{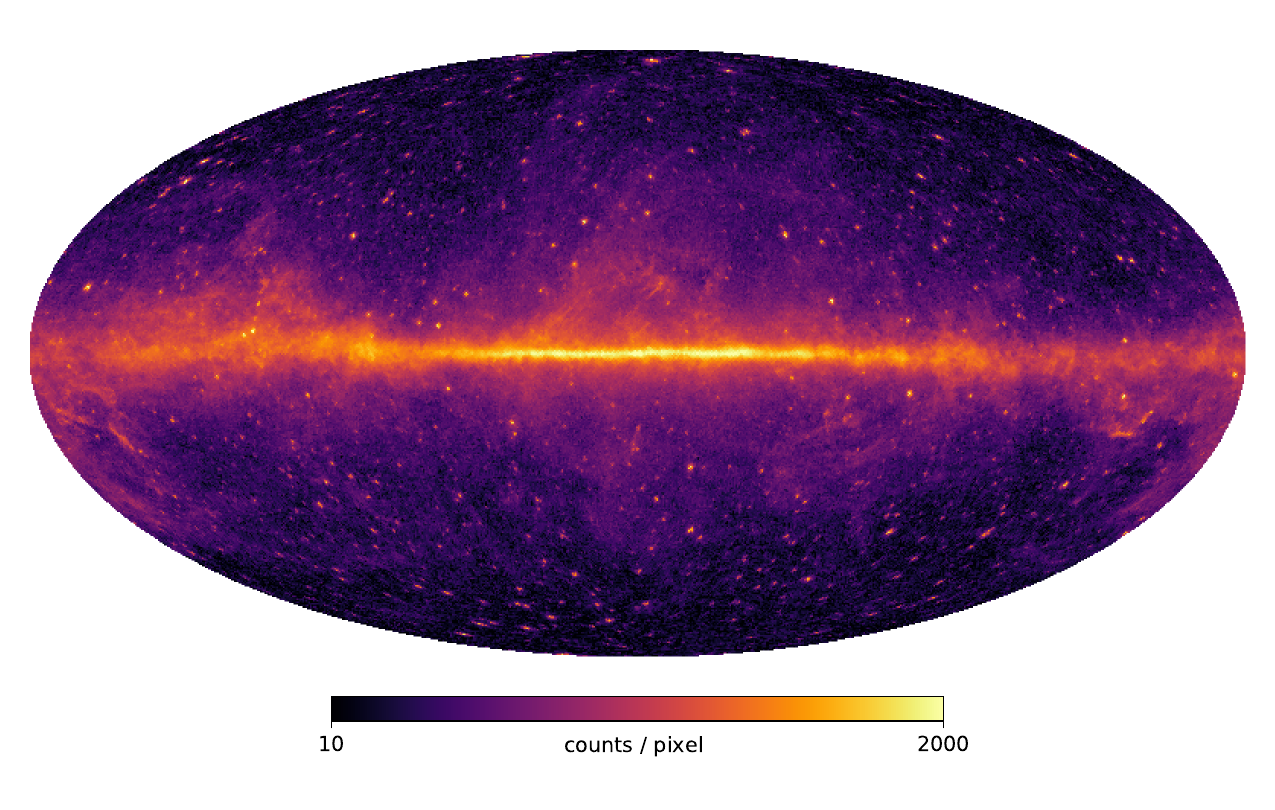}
\includegraphics[scale=0.6]{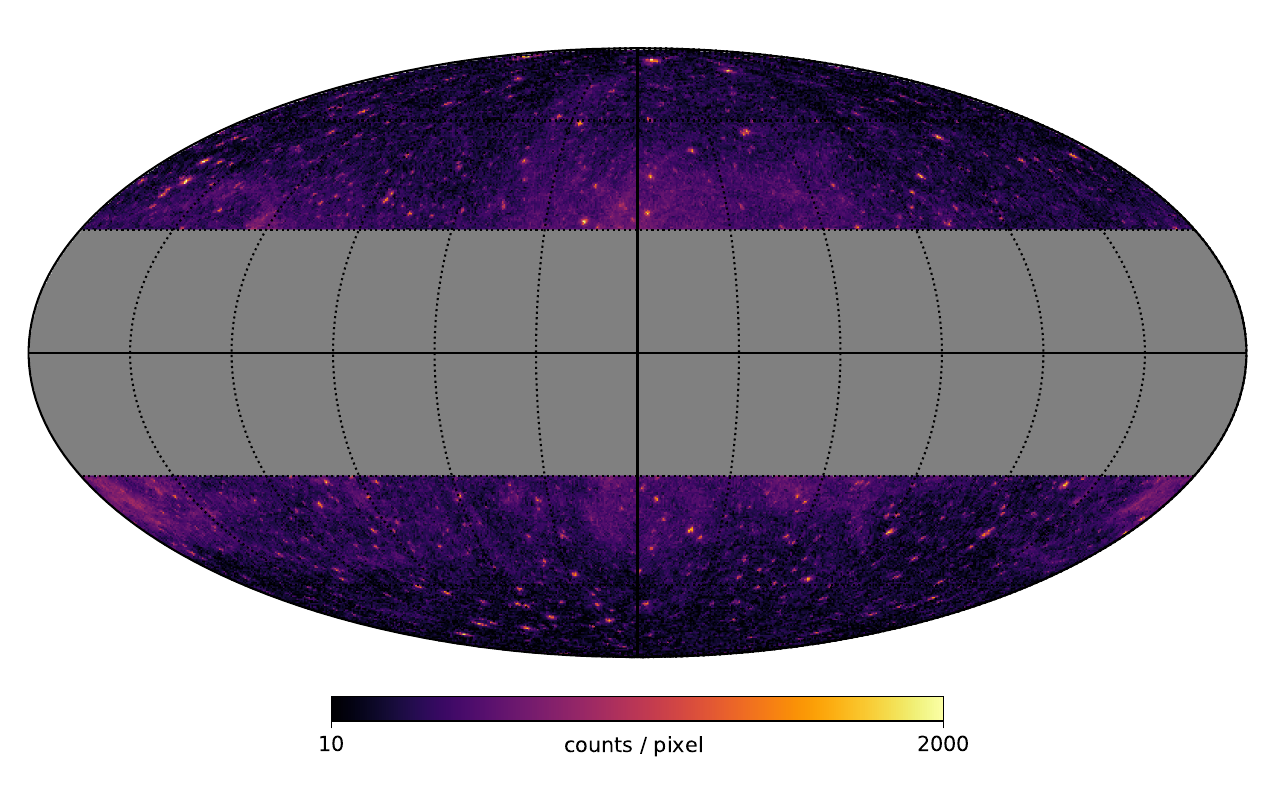}
\caption{Simulated map in units of counts per pixel after  adding Poisson noise from the synthetic model map of Fig. \ref{fig:map-total-counts}, at the same Healpix resolution $N_{\rm side} = 128$ ($n=7$). The lower panel shows the map with a $|b| < 30^\circ$ latitude cut.}
\label{fig:map-effective}
\end{figure}

With a flux map ${\cal M}$ available, we can now produce a synthetic photon-count map ${\cal N}$ as described in Eq. \ref{eq:count}. 
One example is shown in Fig. \ref{fig:map-total-counts}. 
In order to generate maps compatible with a counting experiment like \fermip, we then add Poisson noise to the map performing a Poisson realization of the number of photons in each pixel of the map ${\cal N}$, and we dub this final map as ${\cal N}_{\cal P}$ count map. One example is shown in Fig. \ref{fig:map-effective}, which also shows the same map with the latitude cut applied: these are the type of maps that we use to train and validate the neural network. We have generated in total 1 million maps, 90\% of which have been used for training and 10\% for validation. The maps have been stored as a \verb|TFRecord| dataset\footnote{\verb|https://www.tensorflow.org/tutorials/load_data/tfrecord|} for use with Tensorflow \cite{tensorflow2015-whitepaper}. The maps have been generated in Healpix format with $N_{\rm side} = 128$ resolution (order $n=7$) and then downsampled to $N_{\rm side} = 64$ (order $n=6$) for the training and validation (this resolution corresponds to 49152 pixels). 

The training of the CNN and the subsequent analysis of the \fermi data can be performed equivalently with maps in units of photon-count or in units of flux. Nonetheless, we will work with maps in units of flux, which are easily obtained from the ${\cal N}_{\cal P}$ count maps by using the exposure map and by inverting Eq. (\ref{eq:count}).  The reason to convert the ${\cal N}_{\cal P}$ count maps in units of flux is motivated by the fact that 
count maps contain a spurious large-scale angular dependence
introduced by the exposure map (see Fig. \ref{fig:exposure}) which
might confuse the CNN, while this potential issue is avoided
using flux maps.
Finally, to further ease the learning process of the CNN
we also apply a foreground subtraction procedure,
i.e., to the above flux maps we subtract the original
galactic template flux map.
In this way the CNN can focus on learning the $\dnds$, seeing the exposure and galactic foregrounds only as sources of noise, instead of trying to learn their detailed structure. These foreground-subtracted flux maps are the maps we use to train the CNN. 

\section{Neural network architecture and training}
\label{sec:architechture-and-training}

In this Section we describe the architecture of the CNN we adopt and how data are handled.
The input of the CNN are synthetic foreground-subtracted flux maps, and the output vector ${\hat y}_i = (\dnds|_j, \fiso)$ contains the reconstructed $\dnds|_j$ in 20 flux bins in {$[5 \cdot 10^{-12}, 1 \cdot 10^{-7}]$} cm$^{-2}$ s$^{-1}$, plus the reconstructed value of $\fiso$. The index $i$, thus, has values $i=1,...,21$.
A possible alternative would be to choose as output
the breaks and the slopes of the $\dnds$, i.e., the same parameters used to model the input $\dnds$. Nonetheless, we found a better stability of the CNN using as output directly the $\dnds$ function as discretized above, and we will thus use this as output. 

As further described below, we will use an approach in which we will let the CNN determine also the errors on the output parameters. The output vector will thus be effectively doubled. Beside the ${\hat y}_i$, we thus have also $\hat{\sigma}_i$ the output vector of the errors.

Since we are dealing with sky maps, we can approach the problem from the point of view of image analysis. However, the image here is in spherical projection. We therefore need to adopt a method able to deal with this situation.

\subsection{Spherical neural networks}
In the last years, the machine learning community has realised the importance of developing neural network architectures on complex topologies. One of the most interesting non-trivial domains is that of a sphere, since it cannot be mapped on a plane without introducing distortion. This is the case we are confronting here, since we are dealing with sky maps.  There are many efforts to implement CNN architectures that can do convolutions on the sphere \cite{DBLP:journals/corr/abs-1708-00919,DBLP:journals/corr/abs-1902-04615,DBLP:journals/corr/abs-2012-15000,nnhealpix,mapped-convolutions}, and while there have been many proposals, there is not yet agreement on a standard algorithm.

In dealing with information extraction from spherical images, one has to cope with a set of issues: 
i) the algorithm needs to properly deal with the topology of the sphere without distorting the image or losing information;
ii) the algorithm has to be computationally efficient;
iii) it would be desirable to be able to use the same neural network architectures which already work for flat images, which have been proven to be very efficient and effective.

In fact, a full spherical architecture is usually slower than a flat architecture when applied to images of the same size, and it often happens that it is not possible to adopt pre-existing models without re-implementing them {\it ad-hoc}. 
On the other hand, the adoption of a flat-image architecture either introduces spherical distortions, due to the mapping of a spherical image on a plane, or may loose some of the large scale information contained in the image.

In this work, we  implemented a fast and reliable spherical architecture on the Healpix pixelisation of the sphere, taking inspiration from the idea proposed in \cite{mapped-convolutions} for the icosahedron. This method proved to be reliable when the information we need to retrieve from the maps is a small-scale effect, like in our case the distribution of sources, which are point-like (although smeared by the PSF) and isotropically distributed.

The method is based on the fact that, in the Healpix pixelisation scheme, each pixel has equal area and contains the information of the underlying field (in our case, the photon counts) averaged over that pixel. As discussed above, we train our neural network on maps of order $n=6$ ($N_{\rm side}=64$), which contain 49152 pixels. Considering instead the Healpix sphere with base pixelisation  of order $n=0$ ($N_{\rm side}=1$), we can subdivide the map into 12 equal-area patches, each of which containing 4096 pixels. Each patch is an independent realisation of the underlying isotropic field and can be mapped into a 2D flat image. We call this parsing algorithm \verb|map2patch|. Notice that this is not a re-sampling of the map: it's just a subdivision of the spherical image into 12 patches, in a way that preserves the area and content of each pixel. This allows us to use standard convolutional network techniques that work efficiently on flat images, while preserving the whole available statistics of the full map.

Convolution is performed on each patch separately. In order to be able to extract information from each patch in parallel, we define 3D convolutions, where the first two dimensions are the spatial dimensions, and the third dimension specifies the patch. In order to force the neural network to learn from each individual sky patch, we employ convolution filters of size $(N,N,1)$, where $N$ is the dimension of the filter. By doing so, we effectively apply 2D filters to the whole sky. Note that by employing such filter size, the output of each convolution will still have 12 separate images on the third axis, associated to each sky patch, while the number of channels will vary according to the number of filters employed. This technique will thus naturally scale for images with different color channels, such as RGB images, or spherical images for which several channels are available, as is the case where multiple energy bands are considered at once.

{  We stress that with our method we partially loose large-scale information and information stored at the border of the patches.
Thus our algorithm would not be suitable in the case one is interested in recovering large-scale features.
In our case, however, we are interested in exploring information which is  statistically isotropic and/or lies at scales much smaller then the size of a patch. In this case our architecture is certainly viable and extremely efficient.}
Also, since  we  build our patch on the pixels of the Healpix pixelisation scheme, there is no spherical distortion in our images, unlike other architectures which employ the euclidean projection of the sphere.

This approach is extremely fast, since it leverages on optimised 3D convolutions. Furthermore the sphere-to-patch operation is performed asynchronously by the CPU while the accelerator is training the model, allowing for virtually no bottlenecks in the processing pipeline.

In order to cross-check our method, we also implemented the full-sky truly-spherical convolutions of \cite{nnhealpix}, obtaining very similar results. {  In  Appendix \ref{sec:full-spherical-nn}, we report some examples of the results and checks performed with a fully spherical convolution model.} The two methods are therefore interchangeable for the type of information we want extract from the \fermi maps: however, our implementation allowed us to gain more than a factor of 10 in time of execution of the whole pipeline of training and validation. This convergence of results with the fully spherical method of \cite{nnhealpix} gives us confidence on the reliability of our model, and the speed gain sets the basis for a future efficient extension of our analysis to several energy bins in the full \fermi energy range, in order to determine also the energy dependence of the $\dnds$  of unresolved sources.

\subsection{Data pre-processing}
\label{sec:preprocessing}

The maps are constructed with the procedure outlined in Section \ref{sec:generation}. As explained more in detail there, even though we apply a latitude cut to the maps, in order to further reduce the impact of the galactic foreground (and as it is customary in analyses of the extragalactic gamma-ray background), we subtract the galactic foreground (normalized by the $A_{\rm gal}$ constant) from the input maps. This will also facilitate to some extent the training of the CNN, whose goal is to reconstruct the source-count distribution. Photon-counts maps are converted in units of flux, by using the exposure map. In order to reduce the variability of the inputs seen by the CNN and stabilize its behaviour, we take the logarithm of the flux values in the pixels and then map them in $[-1,1]$ range. 
Finally, each map is parsed into 12 patches using the \verb|map2patch| algorithm discussed in the previous Sections.

Given the approximate $S^{-2}$ behavior of the $\dnds$, for the output vector, we reconstruct $y_i = S^2 \dnds|_i \times 10^{11}$ cm$^{2}$ s deg$^{2}$ ($i=1,20$) and $y_{21}=\fiso \cdot  10^{7}$ cm$^{2}$ s sr, in order to work with (pure) numbers of order unity. We adopt this strategy in order to avoid that the mean square error could favour the optimisation of any flux bin over another, making the deviations more even across the range of $S$.

\subsection{Network architecture}
In recent years, we have seen the rise of many convolutional neural network models for computer vision tasks \cite{CNNreview1, CNNreview2, CNNreview3}.
Modern convolutional neural network architectures can have a very large numbers of internal parameters. This is due to the complex links between each neural unit, which allows for these models to be highly expressive and effectively behave as universal approximators for many kinds of tasks \cite{HORNIK1989359}. Unfortunately this comes with a downside: these models are prone to fitting too well the training dataset, which would lead to poorer performance on new data. In the field of Machine Learning, this problem is known as "overfitting" \cite{overfitting}. In order to prevent overfitting, 
a possible solution is to train the model on a big enough dataset.
This will render impossible for the NN to learn exactly the representation of the training set, as well as showing a richer feature space, which will result in better performance on any additional sample for which we may want to get a prediction, like the real case scenario with the Fermi photon counts map. Another solution is to apply regularisation layers which force the NN to learn a more general representation of the data, by promoting several metrics. Some of the most common regularisation techniques include L1 and L2 regularisation, batch normalisation, and dropout \cite{dropout, batch-normalization, regularisation1, regularisation2}.

We choose as our convolutional neural network architecture the EfficientNet V2M model \cite{efficient-net-v2}. This model has been proven to be faster to train then other models for computer vision tasks, while retaining the same expressivity and predictive power. 
Furthermore, its low memory footprint helps in training with more data at the same time (larger batch sizes), which in turn reduces the risk of overfitting.
In this work, following the original implementation of EfficientNet, we will adopt batch normalisation layers after each convolution.

We additionally employ a fairly new trainable implementation of dropout, called concrete dropout \cite{concrete-dropout}, before each convolution layer.
Dropout \cite{dropout} has been proven to be an efficient way to reduce overfitting in a model, by randomly turning off neurons or connections inside a model. Unfortunately, the dropout probability for each dropout layer has to be either hand picked or optimised through grid search, which is practically unfeasible for very large models. If the dropout probability is too small, the regularisation effect will not be very significant, while on the other hand too high a dropout probability will result in slower training and worse overall performance (underfitting). 
We therefore adopt the implementation of dropout as a concrete Bernoulli distribution \cite{concrete-distribution}, which allows the self-learning of the optimal dropout probability, effectively saving a conspicuous amount of time while training and ensuring that the model will be less prone to overfitting.

\subsection{Bayesian error estimation and cost function}
\label{sec:bayesian-error}

In order to correctly estimate the error on a prediction, it is necessary to correctly estimate both the systematic error (sometimes called epistemic error \cite{epistemic-err}) and aleatory error components of a measurement. One way to estimate such components together is through a Bayesian approach, and in particular by defining a Bayesian neural network capable of estimating the posterior distribution of the desired observables.   
It has been shown \cite{mcdropout} 
that a NN with arbitrary depth and  non-linearities, with dropout applied before every weight layer, is mathematically equivalent to an approximation of a Bayesian model. 

In a subsequent analysis \cite{heteroscedastic-error} it has been shown how to estimate the first and second momenta (mean and variance) of the posterior distribution, under the assumption of a Gaussian likelihood of unknown variance. In particular \cite{heteroscedastic-error} shows how to define a neural network capable of determining the appropriate variances of such a multivariate Gaussian.
In our case, training a neural network with such an approach is advantageous, as the neural network will learn to weight less the component of the measurement at low fluxes, which we expect to be more uncertain, which helps in obtaining stable measurements for the test dataset. 

Following \cite{heteroscedastic-error} we define the heteroscedastic Gaussian negative log-likelihood:
\begin{equation}
    {\sl NLL} = \sum_{i=1}^N  \left[\frac{(y_i^{\rm true}-y_i^{\rm pred})^2}{2\sigma_i^2} + \frac{1}{2} \log(\sigma_i^2)  +  \frac{1}{2} \log(2\pi)  \right]
\label{costfunc}    
\end{equation}
where we compare the true $\dnds$ and $\fiso$ (transformed as discussed in Section \ref{sec:preprocessing}) used to generate the simulated map with the prediction of the neural network, and thus $N=20+1$.
For the training of the neural network, we omit the last term of Eq. (\ref{costfunc}) which is an irrelevant constant.
Regarding the errors, since the first term in Eq. (\ref{costfunc}) is a decreasing function of the $\sigma_i$ while the second is increasing, this ensures that there will be a minimum which will provide the  optimal estimate for the $\sigma_i$ themselves.

In principle, the above cost function can easily be generalized to account for the covariance among the different bins and $F_{\rm iso}$. 
 In practice, however, this implies estimating also the off-diagonal terms of a $21 \times 21$ symmetric matrix. Such matrix has to be inferred autonomously from data without any supervision, making the problem very challenging to optimise in this form. We thus refrain from attempting estimating the covariance. For CNN applications where the covariance between a set of parameters has been estimated, see e.g. Ref. \cite{Mishra-Sharma:2021oxe, List:2020mzd}.

\subsection{Training and validation}
\label{sec:NN-training}

\begin{figure}[t]
\centering
\includegraphics[width=0.6\textwidth]{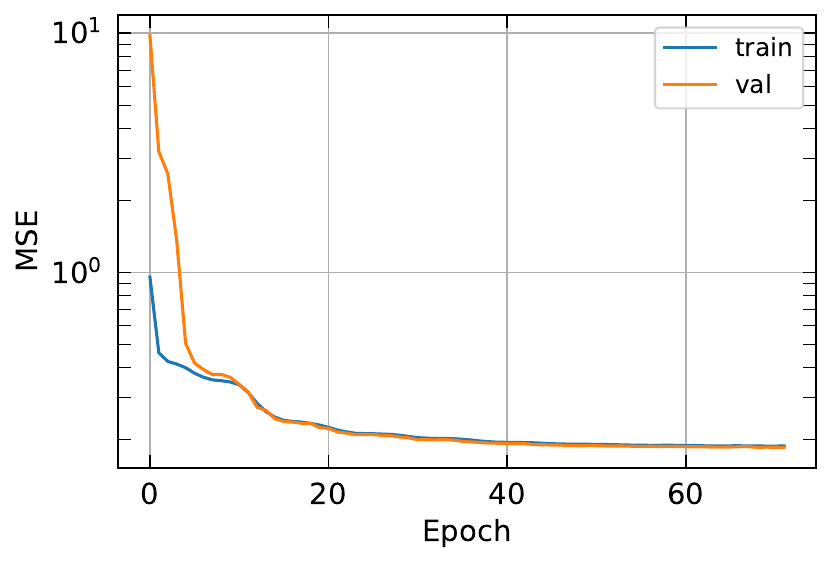}
\caption{ Epoch evolution of the MSE for the training (blue line) and validation (orange line) sets.}
\label{fig:mse-training}
\end{figure}

\begin{figure}[t]
\centering
\includegraphics[width=1.0\textwidth]{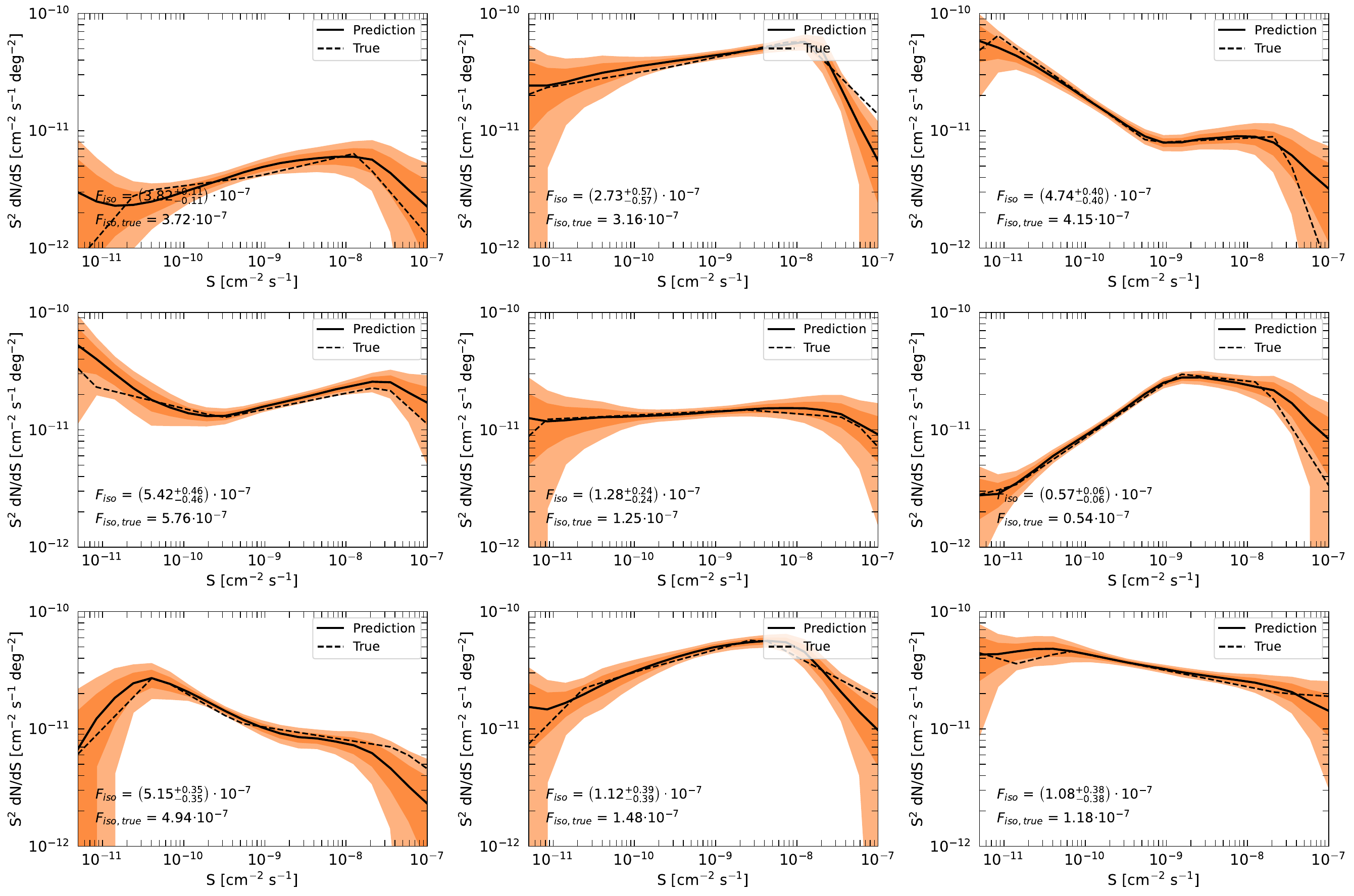}
\caption{Some example $\dnds$ from the validation dataset. Each panel shows the input $\dnds$ and $\fiso$ and the corresponding quantities reconstructed by the CNN. The colored bands indicate 1-2 $\sigma$ uncertainty regions. See text for more details. }
\label{fig:random-maps}
\end{figure}

We train our model using a Tensor Processing Unit (TPU) v3, kindly provided by the Google$^{\rm TM}$ TPU Research Cloud. We adopt the Adam optimizer. We employ a cyclical triangular learning rate \cite{triangular} to avoid falling into local minima of the parameter space. We adopt a batch size equal to $128\times 8=1024$ to fully exploit the computational capabilities of the TPU. We train the model for 72 epochs, for a total of 9 triangular learning rate cycles, with learning rate ranging from $10^{-6}$ to $10^{-3}$. 
As already anticipated, we have generated 1 million maps, 90\% of which have been employed in the training and 10\% for validation.

To test the convergence of the model we evaluate the mean square error (MSE)  defined as $MSE=\sum_{i=1}^N (y_i^{\rm true}-y_i^{\rm pred})^2$. 
In Fig. \ref{fig:mse-training} we plot the MSE as a function of the training epoch for the training and validation dataset. We can see that, starting from epoch 10, the validation and training lines are close to each other, which means that our model is not overfitting the training dataset. We observe that the MSE reaches a plateau at around epoch 60, meaning that further training of our model would not lead to  improvement in performance.

Once the CNN is fully trained, we extensively check the convergence of the reconstructed $\dnds$ and $\fiso$ against their inputs, by using the maps of the validation set. 
Fig. \ref{fig:random-maps} shows a few illustrative examples of $\dnds$, randomly generated according to the priors and parameters intervals listed in Table \ref{tab:prior-dNdS}, together with their reconstructions. 
The dashed black lines show the input $\dnds$ and the black solid lines denote the corresponding distributions reconstructed by the CNN. The  input and reconstructed values of $\fiso$ are also quoted. The colored bands refer to the $1\sigma$ and $2\sigma$ confidence intervals
as estimated by the CNN.
The CNN recovers a $\dnds$ which is compatible with the input. The error bands are large for high fluxes, due to the fact that the number of sources at those values of $S$ is low and therefore a large statistical uncertainty is expected. The error band becomes large again at low fluxes, indicating that the CNN reaches a confusion limit: even though here the number of sources is very large, their faintness makes progressively harder to identify their number and a confusion between faint sources and the average isotropic component $\fiso$ arises, the lower the valus of $S$ becomes. This same behaviour also occurs in the 1point-PDF analysis of \cite{Cuoco-1pdf}. Concerning the value of $\fiso$, the reconstruction is very good, compatible with the input and with a small uncertainty.

\subsection{Frequentist error estimation and Bias} 
\label{sec:bias}

\begin{figure}[t]
\centering
\includegraphics[width=1.0\textwidth]{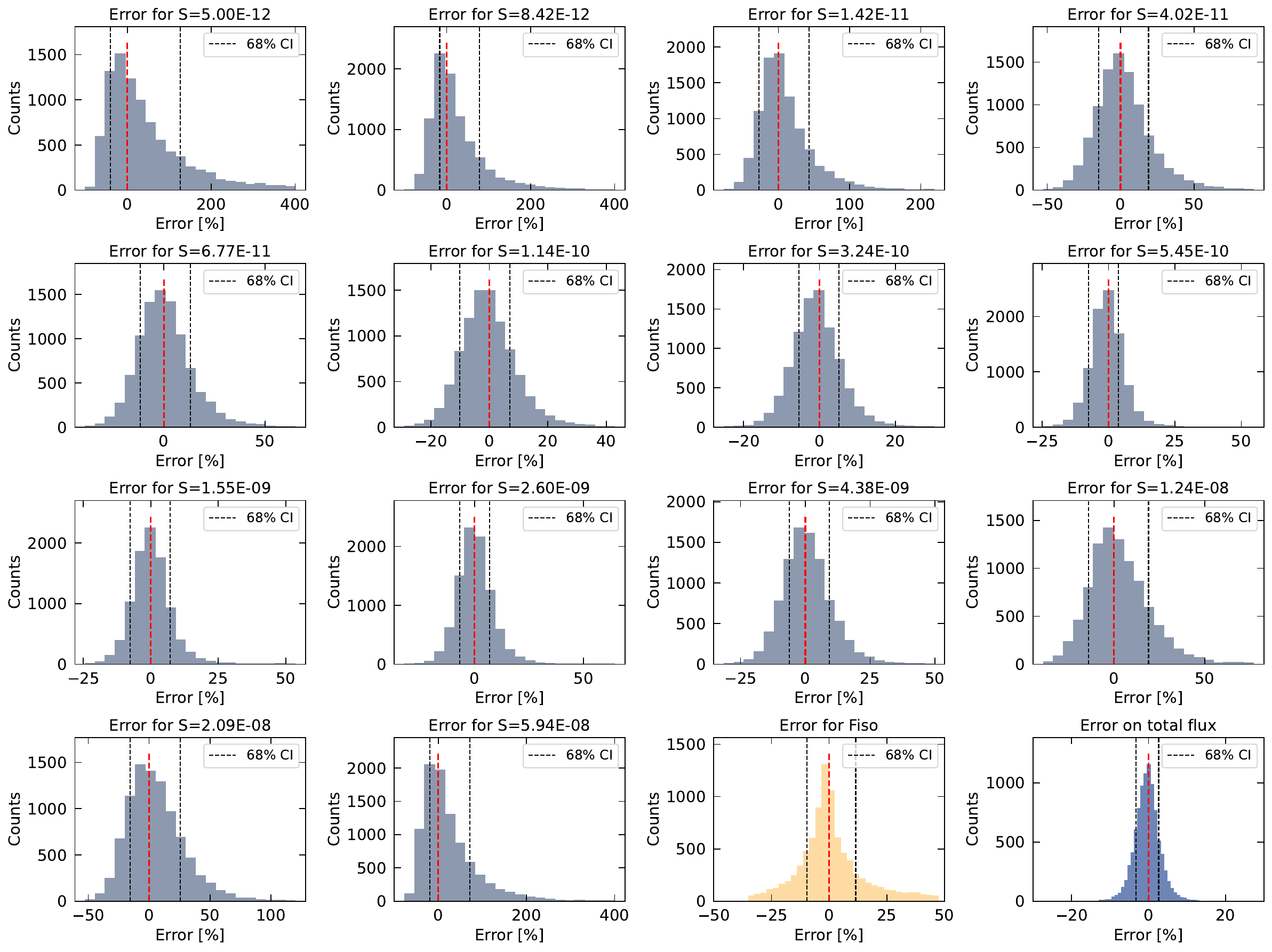}
\caption{ Histograms of the percentage deviations between the true and reconstructed $\dnds$ for 14 values of $S$ across its interval of interest, and for $\fiso$. The bottom right panel shows the same, but for a derived quantity, namely the total flux. The dotted vertical black lines indicate the 68\% confidence region.}
\label{fig:bias-histogram}
\end{figure}

\begin{figure}[t]
\centering
\includegraphics[width=0.8\textwidth]{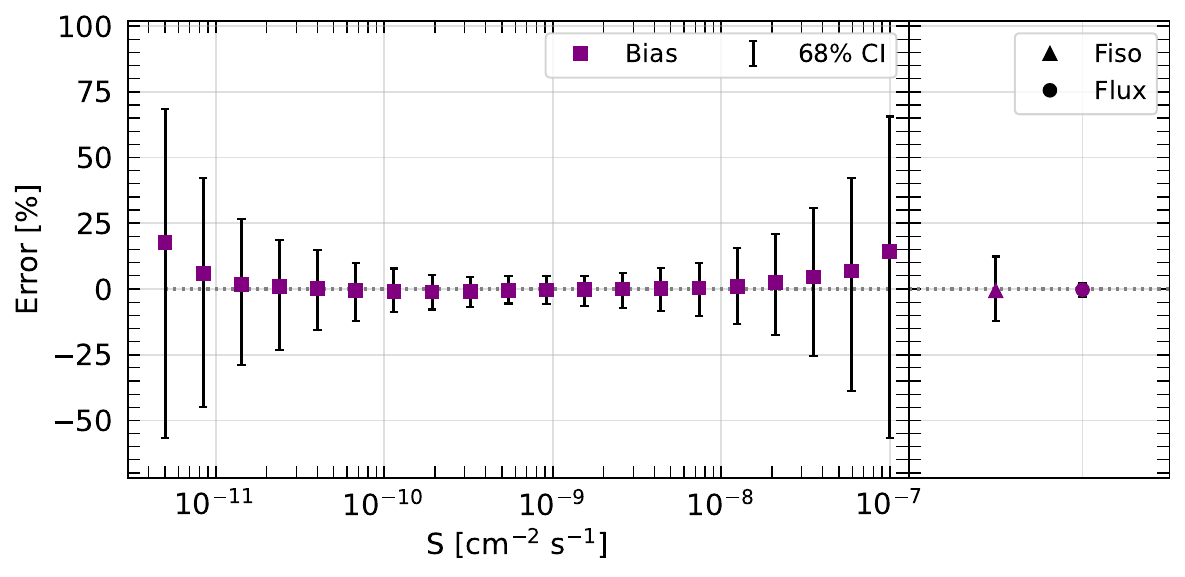}
\caption{68\% error for the reconstructed $\dnds$ as function of $S$, for $\fiso$ and for the total flux. Also shown is the bias given by the median of the histogram for each $S$ bin (squares).
}
\label{fig:bias-plot}
\end{figure}

Besides the Bayesian error automatically provided as output by the CNN, we also estimate the error in an alternative frequentist-like way, in order to provide a cross-check. This second method gives as a further advantage also the possibility to estimate the amount of bias on the reconstructed $\dnds$.
To this aim, we have built, for each of the 20 bins in flux and for $\fiso$, the histogram of the deviations between the true and reconstructed $\dnds$  and $\fiso$. The results for 14 values of $S$ across its interval of interest, and for $\fiso$, are shown in Fig. \ref{fig:bias-histogram} (the remaining histograms are similar in content and are not shown here for economy of space). The plot shows the distribution of models which exhibit a specific relative deviation (expressed as percentage of the true value). By using these
distributions of the deviations (normalized to unity), we derived the $1\sigma$ and $2\sigma$ confidence level intervals on the reconstructed $\dnds$ and $\fiso$. As anticipated in Section \ref{sec:NN-training}, at low and large fluxes the distributions are wide, implying a poorer ability to reconstruct the underlying physical models, while in the flux interval $(1 \cdot 10^{-11}, 2 \cdot 10^{-8})$ cm$^{-2}$ s$^{-1}$ the $1\sigma$ error is below 30\%, with an error below 20\% for $S$ in $(1 \cdot 10^{-10}, 1 \cdot 10^{-8})$ cm$^{-2}$ s$^{-1}$. The $1\sigma$ error on $\fiso$ is also of the order of 20\%. The last panel of Fig. \ref{fig:bias-histogram} shows the histogram for the reconstructed value of the total photon flux: this is a sanity check and shows that the CNN correctly reproduces the normalization of the maps in terms of total counting rate. The uncertainty on the total flux is 5\%. The $1\sigma$ (non symmetric) intervals for all the 20 flux bins are also reported in Fig. \ref{fig:bias-plot}.
A more direct comparison of the Bayesian and frequentist errors is provided in the results section.

The same histograms of Fig. \ref{fig:bias-histogram} allow us also to check whether the reconstructed values are biased. We define the bias as the deviation between the median of the distribution and the input value. The values of the bias are shown, for each of the 20 values of flux, in Fig. \ref{fig:bias-plot} as squares. We see that for all fluxes, the bias is very close to zero, which implies that the reconstructed values not only are determined with good precision (since the error bar is small), but also without a  bias toward larger or lower values. The only exception is a positive bias for very low fluxes, where the CNN meets its confusion limit, and for very large fluxes, where the low source-counts statistics produces a small overestimate of the underlying $\dnds$.
In both cases, however, the bias is small compared to the overall error.

\section{Results}
\label{sec:results}

\begin{figure}[t]
\centering
\includegraphics[width=0.9\textwidth]
{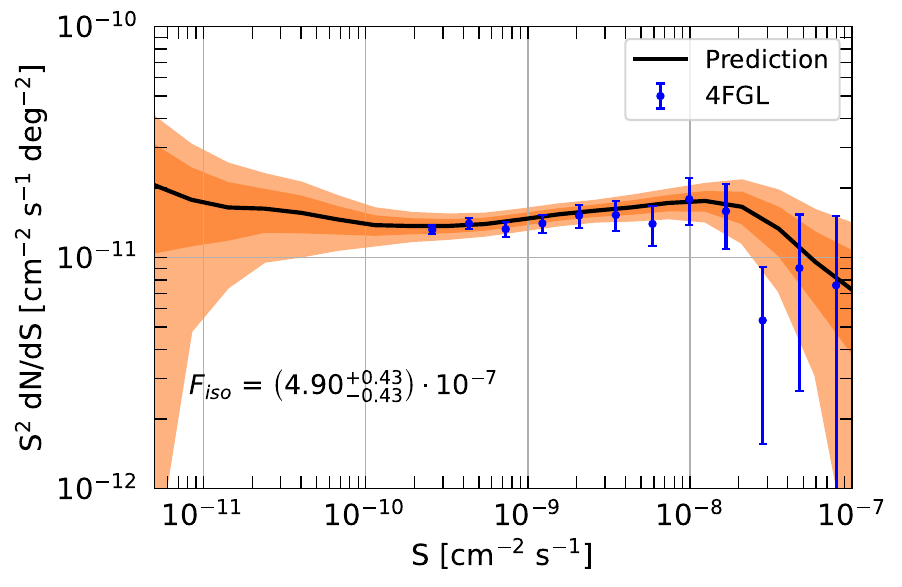}
\caption{Reconstructed $\dnds$ and $\fiso$  and their Bayesian errors, when the trained CNN is applied to the \fermi map. This is the main baseline result of the analysis.}
\label{fig:prediction-fermi-12y}
\end{figure}

After the training and validation processes described in the previous Section, we have applied the fully trained CNN to the 14 years \fermi data. The result of our baseline analysis, which adopts the \verb|gll_iem_v07| galactic foreground and a latitude cut of $|b|<30^\circ$, is shown in Fig. \ref{fig:prediction-fermi-12y}. The blue line is the reconstructed $\dnds$ and the error bands are the Bayesian errors obtained as described in Section \ref{sec:bayesian-error}. The blue points show the $\dnds$ of the resolved sources, obtained from the 4FGL-DR3 catalog. We can see that in the resolved limit, the CNN reconstructs a $\dnds$ fully compatible with the one derived from the catalog, but then extends it to the unresolved regime with a behaviour compatible with $\dnds \sim S^{-2}$ down to the smallest flux considered of $S = 5 \cdot 10^{-12}$ cm$^{-2}$ s$^{-1}$.
Below $S \sim  10^{-10}$ cm$^{-2}$ s$^{-1}$ the $\dnds$ exhibits a slight turn-up, although the feature has a very low statistical significance.

\begin{figure}[t]
\centering
\includegraphics[width=0.9\textwidth]{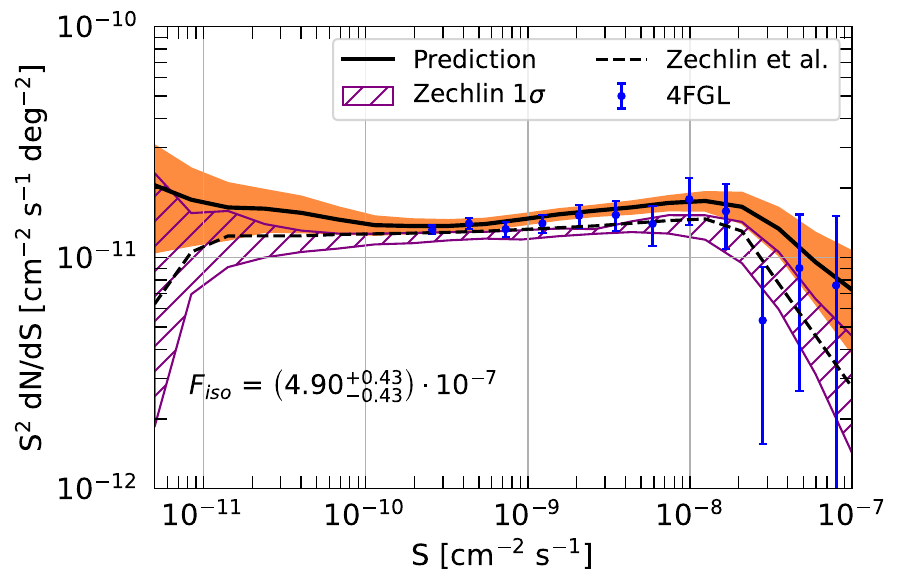}
\caption{Same as Fig. \ref{fig:prediction-fermi-12y}, including a comparison with the result of Ref. \cite{Cuoco-1pdf}.}
\label{fig:prediction-fermi-12y-vs-zechlin}
\end{figure}

The result obtained here is also compatible with the one obtained in \cite{Cuoco-1pdf}, as shown in Fig. \ref{fig:prediction-fermi-12y-vs-zechlin}. The difference between the two analyses, apart from the technique adopted to extract the $\dnds$, stands in the fact that here we update the data set to 14 years of \fermi data collection and to improved detector response functions, as compared to \cite{Cuoco-1pdf}. Another difference is that in \cite{Cuoco-1pdf} the  $\dnds$ was reconstructed with a broken power-law, while here we are allowing for a more flexible functional dependence (we are using 20 points in the flux interval shown in Fig. \ref{fig:prediction-fermi-12y}). The result is nevertheless quite compatible with a power law behaviour with $\dnds \sim S^{-2}$ for mid-values of $S$, $\dnds \sim S^{-3}$ for $S > 10^{-8}$ cm$^{-2}$ s$^{-1}$ and a slight decrease in the power-law index for fluxes $S < 10^{-10}$ cm$^{-2}$ s$^{-1}$.

\begin{figure}[t]
\centering
\includegraphics[width=0.9\textwidth]{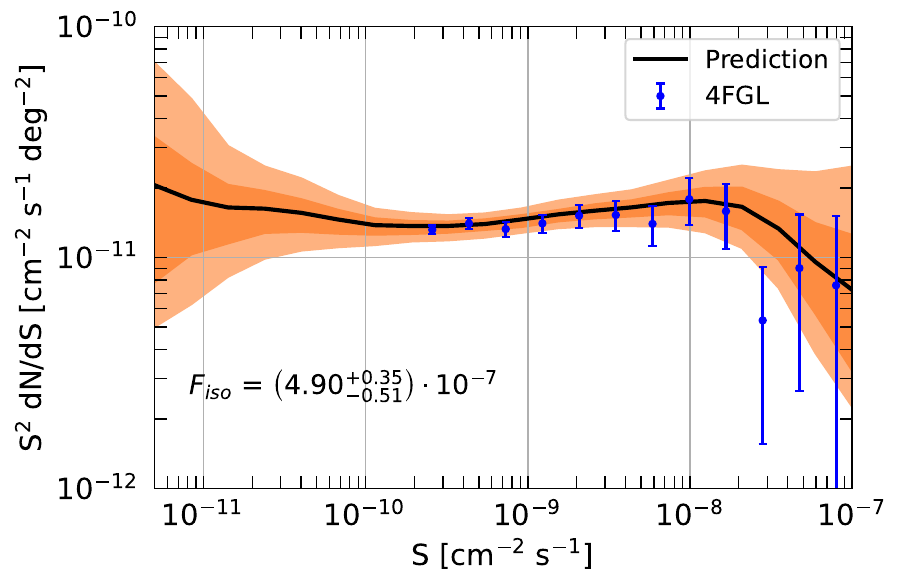}
\caption{Same as in Fig. \ref{fig:prediction-fermi-12y}, but with errors estimated through a frequentist approach.}
\label{fig:prediction-fermi-12y-bayesian}
\end{figure}

In Fig. \ref{fig:prediction-fermi-12y-bayesian} we show the measurement for the Fermi map employing the frequestist estimation of the error. It can be seen that the error is very similar to the Bayesian one, with some minimal difference at low fluxes. This provides confidence that the estimated error is robust.

The CNN provides also the value $\fiso=4.90 \pm 0.43 \cdot 10^{-7}$ cm$^{-2}$ s$^{-1}$ sr$^{-1}$, 
which can be compared with the value $F'_{\rm iso}=4.91 \pm 0.04 \cdot 10^{-7}$ cm$^{-2}$ s$^{-1}$ sr$^{-1}$ reported in Table \ref{tab:fore-parameters}.
We remind that $F'_{\rm iso}$ is the sum of $\fiso$ and of the contribution of sources below the 4FGL catalog threshold. Indeed, if we  calculate the contribution of sources using our best fit $\dnds$ integrated in the range $[5 \cdot 10^{-12}, 1 \cdot 10^{-10}]$ cm$^{-2}$ s$^{-1}$, the upper limit being an approximate value for the catalog threshold, and we add $\fiso$  we obtain a value of {$(6.41 \pm 0.61)\cdot 10^{-7}$ cm$^{-2}$ s$^{-1}$} sr$^{-1}$.
This is in slight tension with $F'_{\rm iso}$, although the two agree at the $\sim 2\sigma$ level.

\begin{figure}[t]
\centering
\includegraphics[width=0.9\textwidth]{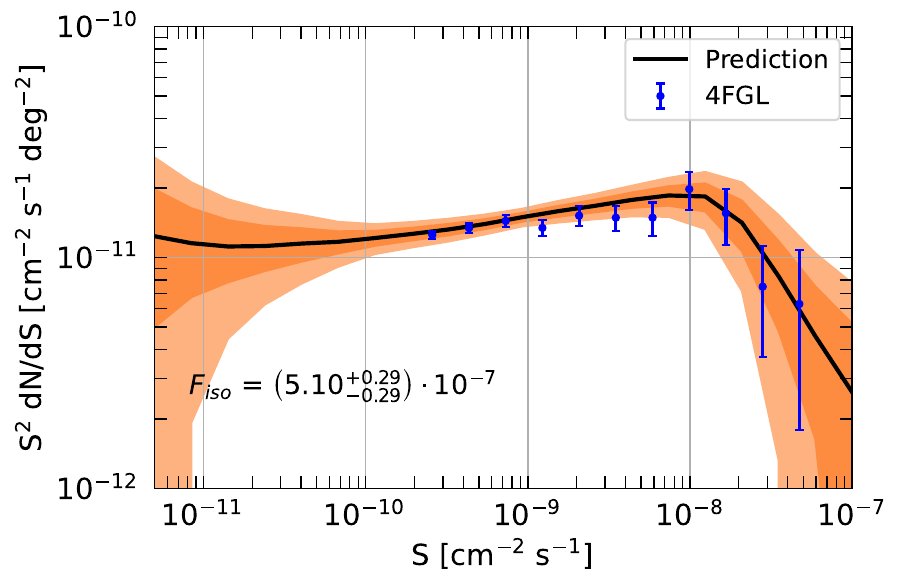}
\includegraphics[width=0.9\textwidth]{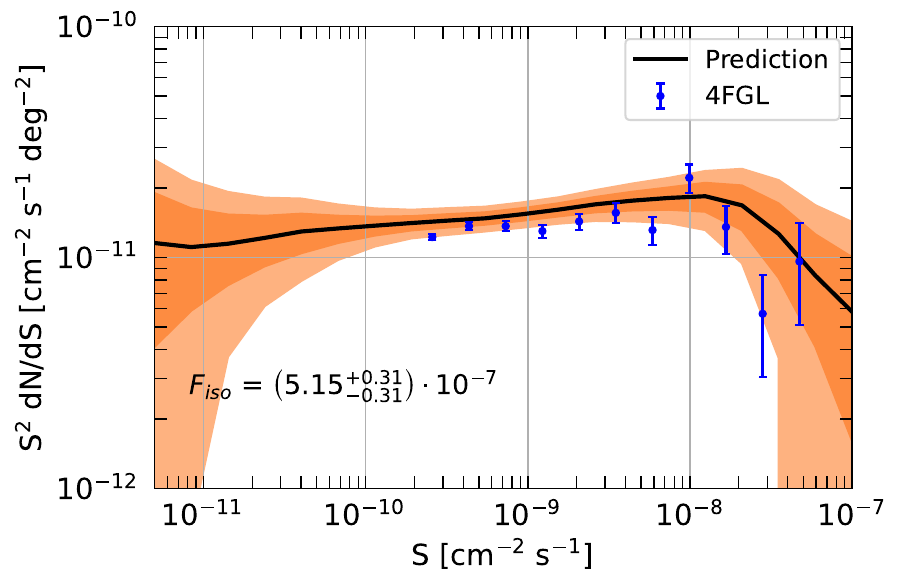}
\caption{Same as Fig. \ref{fig:prediction-fermi-12y} but for a latitude cut of 40$^\circ$ and 50$^\circ$. The blue datapoints of the $\dnds$ of the 4FGL sources have been updated using only the source in the given region of interested. }
\label{fig:prediction-fermi-12y-b4050}
\end{figure}

In order to test the stability of the results, we repeated the analysis by using the same \verb|gll_iem_v07| galactic foreground but with two different latitude cuts: $|b|<40^\circ$ and $|b|<50^\circ$. The positive consequence of a higher latitude cut is a lower residual foreground contamination. However, the amount of available data is reduced. The CNN has been fully re-trained for both of these situations, and the ensuing results when applied to the \fermi data are shown in Fig. \ref{fig:prediction-fermi-12y-b4050}. We notice that the results, both in terms of behaviour and uncertainty estimate, are fully compatible with the results of the baseline analysis. The $\dnds$ exhibits a slight decrease toward lower fluxes as compared to the baseline case, although the size of this effect is not statistically significant. The values of $\fiso$ are also fully compatible among them.

\begin{figure}[t]
\centering
\includegraphics[width=0.9\textwidth]{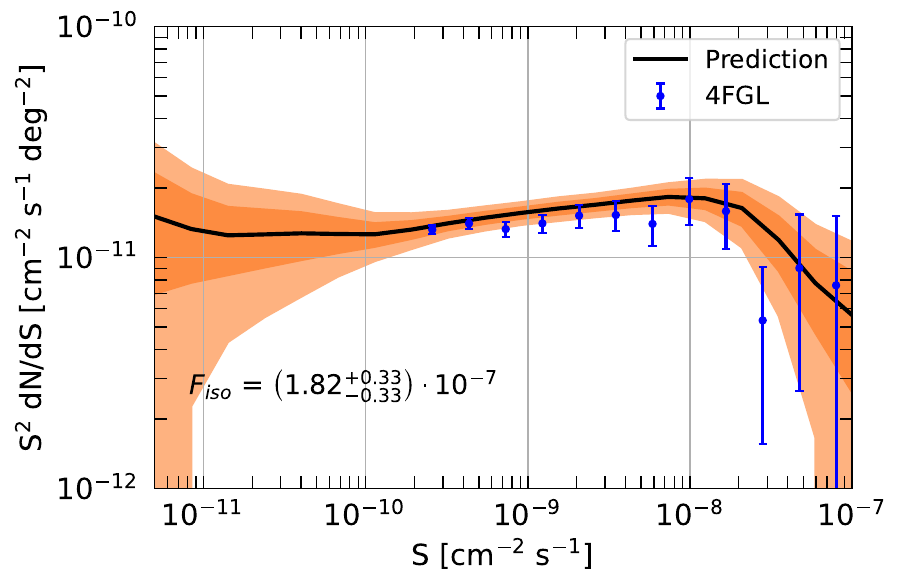}
\caption{Same as Fig. \ref{fig:prediction-fermi-12y} but using a CNN trained using the foreground model v05.}
\label{fig:prediction-fermi-12y-fg-v5}
\end{figure}

A second test of stability has been performed in order to check the impact of galactic foreground modeling. We have performed a new training of the CNN with the alternative template \verb|gll_iem_v05|, and with the $|b|<30^\circ$ latitude cut. The result is shown in Fig. \ref{fig:prediction-fermi-12y-fg-v5}. The reconstructed $\dnds$ matches to a large degree the source-count distributions obtained with \verb|gll_iem_v07|. This reassures us that, even though the galactic foreground modeling introduces a degree of uncertainty, nevertheless the results of the CNN for $\dnds$ are remarkably stable.
The value of $\fiso$ is smaller in this case, equal to
$\fiso=1.82 \pm 0.15 \cdot 10^{-7}$ cm$^{-2}$ s$^{-1}$ sr$^{-1}$, 
which can be compared with the value $F'_{\rm iso}=2.57 \pm 0.06 \cdot 10^{-7}$ cm$^{-2}$ s$^{-1}$ sr$^{-1}$ reported in table \ref{tab:fore-parameters}.
Adding to $\fiso$ the integral of the contribution of sources in the interval in the range $[5 \cdot 10^{-12}, 1 \cdot 10^{-10}]$ cm$^{-2}$ s$^{-1}$ using the obtained $\dnds$, we obtain a total flux of { $(2.92 \pm 0.54)\cdot 10^{-7}$ cm$^{-2}$ s$^{-1}$} sr$^{-1}$, which is in very good agreement with $F'_{\rm iso}$.

In Appendix \ref{furthertests}, we also discuss the results of further tests that we performed to cross-check the stability of the $\dnds$ derived in this section, and that confirm the  robustness of the result.

\section{Conclusions}

In this paper we have studied the differential source-count distribution of gamma-ray sources below the \fermi  threshold for source identification. We  adopted a convolutional neural network methodology devised to deal with spherical sky-maps and able to extract information related to small scale fluctuations. The technique is based on the EfficientNet V2M model, applied to a patched version of the 2-dimensional spherical sky-map projection of the gamma-ray emission. The neural network has been trained on 900k synthetic maps and validated on 100k additional maps. These synthetic maps have been generated from the flux of random selections of source-counts distributions, to which galactic foreground models and an isotropic component have been added. After training and validation, the neural network has been applied to the \fermi sky-map.

We concentrated our analysis on the photon energy interval $(1,10)$ GeV and to 14 years of data. The main result is shown in Fig. \ref{fig:prediction-fermi-12y}: the reconstructed source count distribution exhibits a $\dnds \sim S^{-2}$ behaviour over almost four orders of magnitude in flux in the range  
$[5 \cdot 10^{-12} , 1 \cdot 10^{-8}]$ cm$^{-2}$ s$^{-1}$, which includes the unresolved regime range 
$[5 \cdot 10^{-12}, 2 \cdot 10^{10}]$.
In the regime where \fermi has sensitivity to individually resolve gamma-ray sources, Fig. \ref{fig:prediction-fermi-12y} shows that the neural network fairly reproduces the observations, giving confidence on its reliability and strength.
While in the unresolved regime the result is in agreement with previous studies performed with different methodologies.
We have also further validated the result performing several test of stability which have confirmed that  our findings is stable and robust.

 The methodology presented here provides a proof of principle for the adoption of a CNN to the reconstruction of the source-count distribution of the extra-galactic sky. 
With respect to other methods to extract the source-count distribution in the unresolved regime, like the 1-point PDF one, the use of a CNN avoids the need to calculate complicated and numerically demanding likelihoods.
 Possible future applications includes the extension to multiple energy ranges and energy correlations, and the investigation of features in the $dN/dS$ which might indicate the presence of exotic components like dark matter.
A further aspect which would be interesting to investigate is to compare different approaches to machine learning on spherical domains. In particular, it might be interesting to employ a graph convolutional neural network as suggested in \cite{DBLP:journals/corr/abs-2012-15000}, in order to preserve spatial relations between distant pixels, and rotational invariance.
Finally, recent developments in the field of simulation based inference, such as our case, suggest that a promising approach to estimate the complex posterior distribution of the objective parameters are the so-called neural likelihood ratio estimation techniques \cite{Durkan2020}, and in particular Truncated Marginal Neural Ratio Estimation \cite{Miller2021arxiv,Miller2021,AnauMontel:2022ppb}. Since these techniques are tailored for simulation based problems, we hypothesise that it would be possible to achieve similar performance to our current neural network architecture with an even simpler structure and possibly less training samples. 

\label{sec:conclusions}

\acknowledgments
 We thank B. Zaldivar for valuable discussions on aspects related to convolutional neural network applications and Bayesian error estimation. We thank M. Negro for valuable insight on \fermi data analysis and data reduction and for the numerical tool Xgam\footnote{\verb|https://github.com/nmik/Xgam|}. 
 We thank G. Zaharijas for serving as LAT internal referee and providing useful comments on the manuscript.  
We acknowledge support from: {\sl Departments of Excellence} grant awarded by the Italian Ministry of Education, University and Research (MIUR); Research grant {\sl The Dark Universe: A Synergic Multimessenger Approach}, Grant No. 2017X7X85K funded by the Italian Ministry of Education, University and Research (MIUR); {\sl Research grant The Anisotropic Dark Universe}, Grant No. CSTO161409, funded by Compagnia di Sanpaolo and University of Torino; Research grant {\sl TAsP (Theoretical Astroparticle Physics)} funded by Istituto Nazionale di Fisica Nucleare (INFN). 
This research was supported with Cloud TPUs from Google's TPU Research Cloud (TRC).

The \textit{Fermi} LAT Collaboration acknowledges generous ongoing support
from a number of agencies and institutes that have supported both the
development and the operation of the LAT as well as scientific data analysis.
These include the National Aeronautics and Space Administration and the
Department of Energy in the United States, the Commissariat \`a l'Energie Atomique
and the Centre National de la Recherche Scientifique / Institut National de Physique
Nucl\'eaire et de Physique des Particules in France, the Agenzia Spaziale Italiana
and the Istituto Nazionale di Fisica Nucleare in Italy, the Ministry of Education,
Culture, Sports, Science and Technology (MEXT), High Energy Accelerator Research
Organization (KEK) and Japan Aerospace Exploration Agency (JAXA) in Japan, and
the K.~A.~Wallenberg Foundation, the Swedish Research Council and the
Swedish National Space Board in Sweden.
 
Additional support for science analysis during the operations phase is gratefully
acknowledged from the Istituto Nazionale di Astrofisica in Italy and the Centre
National d'\'Etudes Spatiales in France. This work performed in part under DOE
Contract DE-AC02-76SF00515.

\clearemptydoublepage
\bibliography{bibliography}

\appendix
\section{Further Tests}
\label{furthertests}

In this appendix we describe further tests that we have performed to cross-check the stability and robustness of the $\dnds$ derived in the main text.

\subsection{Flat $S^2\,\dnds$}

\begin{figure}[t]
\centering
\includegraphics[width=1.0\textwidth]{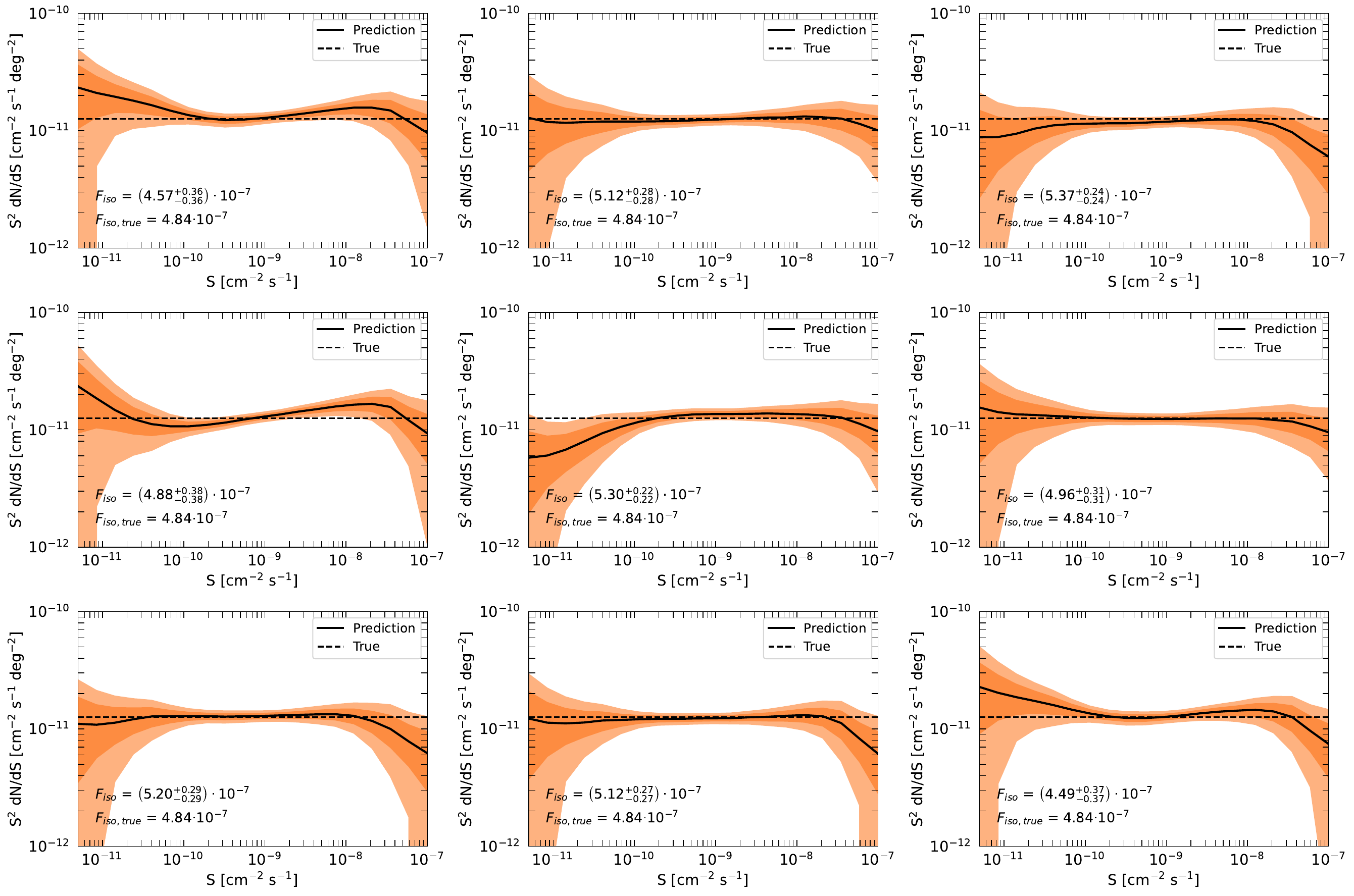}
\caption{  Same as Fig. \ref{fig:random-maps}  but for 6 different random realization of the same flat $S^2\dnds$ and same $F_{\rm iso,true}=4.84 \cdot 10^{-7}$ cm$^{-2}$ s$^{-1}$ sr$^{-1}$.  }
\label{fig:flat-maps}
\end{figure}

\begin{figure}[t]
\centering
\includegraphics[width=1.0\textwidth]{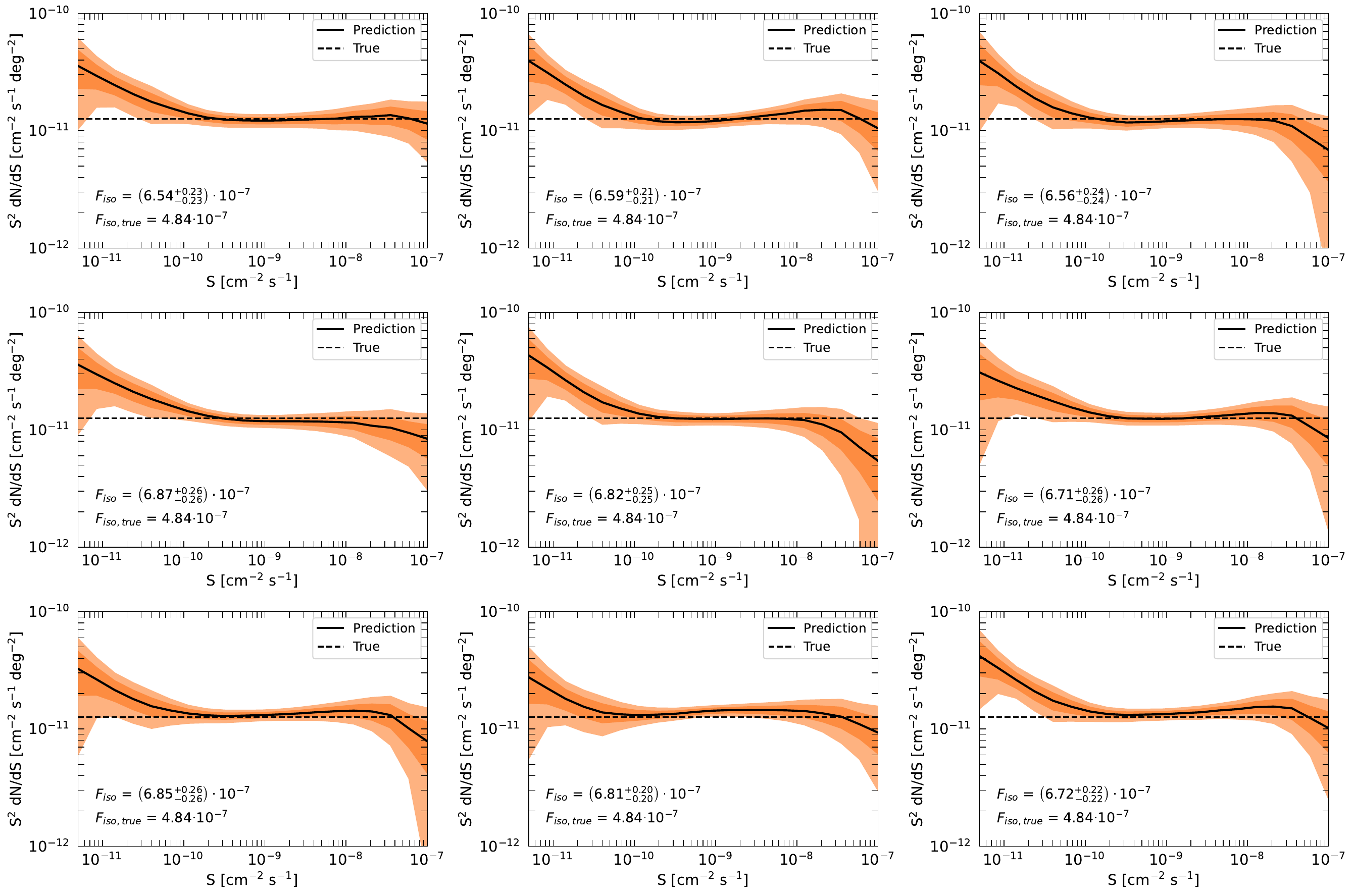}
\caption{
 Same as Fig. \ref{fig:flat-maps} but for synthetic maps generated using version v05 of the foreground model.  }
\label{fig:flat-maps-v5}
\end{figure}

Fig. \ref{fig:flat-maps} shows the stability of the CNN output for a specific repeated $\dnds$ input.
In this case, the CNN, trained on the wide variability of models of Table \ref{tab:prior-dNdS}, is applied to a set of maps, all generated with a specific $\dnds \sim S^{-2}$. This specific choice of input $\dnds$ is meant to test a case which is expected to be similar to the real \fermi case, as was found in \cite{Cuoco-1pdf} for mid-values of $S$. The plots show that the CNN is remarkably consistent in reproducing the correct behaviour. The drop at high fluxes is mostly due to the fact that the CNN has been instructed with maps that contain a dropping $\dnds$ at high fluxes (see Table \ref{tab:prior-dNdS} - the reason for this choice of prior is that sources in the 4FGL catalog exhibit this drop in the source count). For mid-values of $S$, the reconstructed $S^2\dnds$ is quite flat, with a small error band. At low $S$, well below the \fermi threshold and when approaching the confusion limit, some mild deviation occurs, but the true $\dnds$ lies always within the uncertainties. These tests gives us confidence that the CNN should be able to faithfully reconstruct the source distribution of the \fermi sky, when applied to the real data in Section \ref{sec:results}.

As a further test, we show in Fig. \ref{fig:flat-maps-v5} the case where we apply our method to maps which contain a galactic foreground model different from the one used in the training. The CNN has been built (like in our baseline case) with maps generated with the galactic template model \verb|gll_iem_v07|. The same model is used in the foreground subtraction performed in the analysis. On the contrary, for this test we have generated maps  using the alternative \verb|gll_iem_v05| foreground model, and let the above \verb|gll_iem_v07|-trained CNN to analyze them. This has been done in order to verify the resilience of the method on the imperfect knowledge of the galactic foreground. Fig. \ref{fig:flat-maps-v5} shows that the reconstructed $\dnds$ has a level of agreement with the input model comparable to the analogous case of Fig. \ref{fig:flat-maps}. Some larger deviation is present at very low fluxes, even though the reconstructed and input models are compatible within $1\sigma$.

\subsection{UltraCleanVeto Selection}

\begin{figure}[t]
\centering
\includegraphics[width=0.9\textwidth]{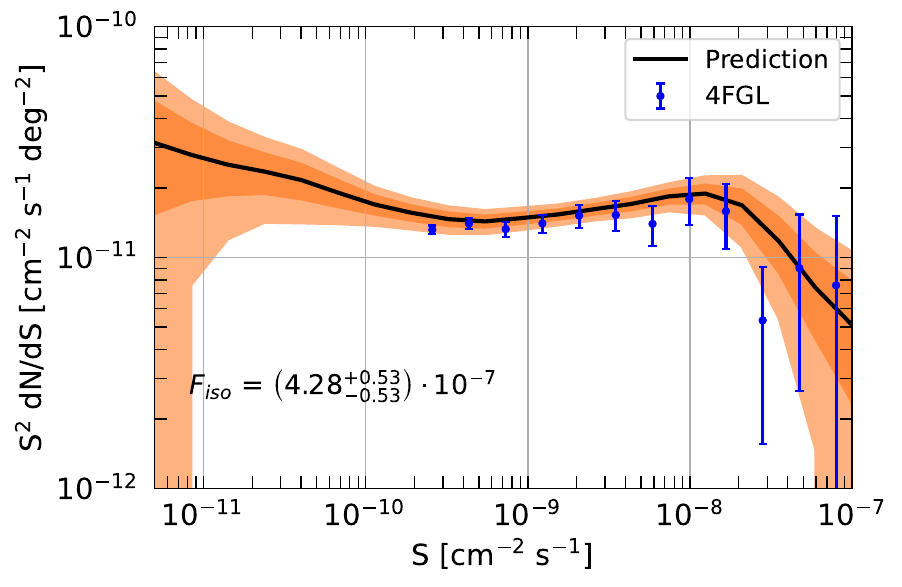}
\caption{Same as Fig. \ref{fig:prediction-fermi-12y} but  for the \texttt{ULTRACLEANVETO}  \fermi data selection. }
\label{fig:prediction-fermi-12y-ucv}
\end{figure}

In this section we further test the robustness of the results to the \fermi data selection. In particular, while for the main results we used the \texttt{SOURCEVETO} event class selection, here we test the \texttt{ULTRACLEANVETO} selection. This event class has more stringent cuts with respct to the \texttt{SOURCEVETO} class, and it thus contains a lower residual charged  cosmic-ray background, at the prize of a lower effective area (by about 15\%). Except for the different data selection all the rest of the analysis is performed in the same identical way as for the \texttt{SOURCEVETO} case. The resulting $\dnds$ is shown in Fig. \ref{fig:prediction-fermi-12y-ucv} and it can be seen that it is compatible with the main results, while, as expected, the value of $\fiso$ is a bit lower.

\subsection{Multipole Analysis}

Since foreground modeling can be an important source of systematic uncertainty,  we investigate this potential issue performing a further test. 
As discussed above, the signal we try to extract from the gamma ray maps, is a small-scale effect, since it is due to a distribution of point sources. To a large degree, this point sources are isotropically distributed in the sky and contribute to the fluctuations of this isotropic field at small angular scales. On the contrary, the galactic foreground is quite diffuse, as can be seen in Fig. \ref{fig:galactic-foreground}, and therefore contributes mainly to large-scale anisotropies. 
In order to further remove from the data the possible large-scale residual component of the galactic foreground, which might still be present after imperfect foreground removal performed with the use of the templates, we adopt a method based on the transformation of the flux to harmonic space and the removal from the maps of the low-multipoles (i.e., large angular scales) contribution. 
We stress that this procedure is applied only to the \fermi data map, and  not to the synthetic maps. For the latter, the foreground subtracted is the same used to generate the maps, so foreground subtraction is, by definition, ``perfect''. 
The procedure follows these steps:

\begin{itemize}
\item We start with the foreground-subtracted flux map, to which we apply the $|b|<30^\circ$ latitude cut and a $2^\circ$ mask around each of the bright sources of the 4FGL catalog (which would otherwise dominate the angular power spectrum).
We first determine and remove the monopole ($l=0$) and dipole ($l=1$) components from the masked map (for this, we use the \verb|remove_dipole| routine from the Healpy library). In this way, after the subtraction procedure outlined below, the monopole (i.e. the total flux) and dipole information of the \fermi map are retained.

\item  This map is then subject to harmonic space decomposition:
\begin{equation}
{\cal M}_{\rm masked}(\theta,\phi) = \sum_{l=0}^{\infty} \sum_{m=-l}^{l} a_{lm} Y_{lm}(\theta,\phi)
\label{eq:multipole}
\end{equation}
where $\theta$ and $\phi$ are angles on the sphere, $Y_{lm}(\theta,\phi)$ the spherical harmonics and the harmonic coefficients $a_{lm}$ completely encode the same information present in the map ${\cal M}_{\rm masked}$  (notice that, even though in Eq. (\ref{eq:multipole}) we formally decompose over all multipoles, the monopole and dipole amplitude are vanishing, since they have been removed from  ${\cal M}_{\rm masked}$);

\item The harmonic coefficients $a_{lm}$ are determined with the \verb|map2alm| Healpy function. The ones that are retained for the foreground cleaning are those referring to multipole values $l < l_{\rm max}$, where $l_{\rm max}$ will be varied and the dependence of the results with $l_{\rm max}$ will be studied (at this level, the monopole and dipole are not present, because of the above point);

\item By using the $a_{lm}$ corresponding to multipoles $l < l_{\rm max}$, we use Eq. (\ref{eq:multipole}) to construct a map ${\cal M}_{\rm residual}$ which contains only the large-scale component. This is performed with \verb|alm2map| Healpy function.

\item We subtract the residual map from the original map, thus obtaining a map where the large scale residual components have been removed. The size of the scale is determined by $l < l_{\rm max}$.

\end{itemize}

\begin{figure}[t]
\centering
\includegraphics[width=1.0\textwidth]{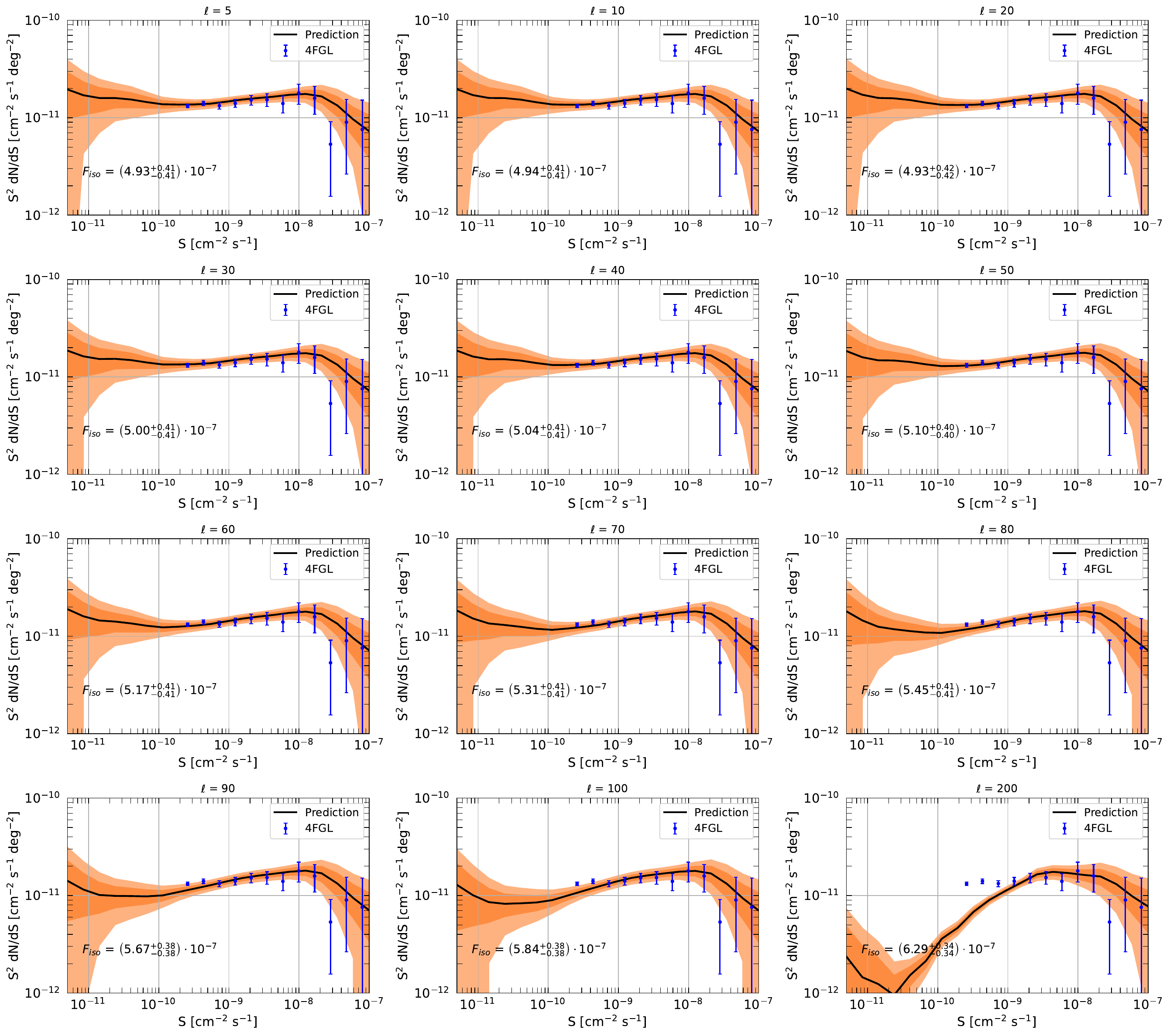}
\caption{Same as Fig. \ref{fig:prediction-fermi-12y} but where the \fermi map given as input to the CNN has been processed with the improved foreground cleaning procedure based on multipole decomposition described in the text. Each panel refer to a different maximal multipole $\ell_{max}$ used for the cleaning.  }
\label{fig:multipole-cleaningv7}
\end{figure}

With this algorithm we obtain maps where the large scale residual features associated to imperfect  galactic foreground subtraction are further removed. The results we obtain on the reconstructed $\dnds$ for the \fermi data (after this ``multipole cleaning") are shown in Fig. \ref{fig:multipole-cleaningv7}, for different values of $l_{\rm max}$ and for the \verb|gll_iem_v07| template. By comparing Fig. \ref{fig:multipole-cleaningv7} with Fig. \ref{fig:prediction-fermi-12y}, which refers to the same foreground template and latitude cut, we can see that the results are extremely stable up to $l_{\rm max} \sim 100$. This means that our baseline analysis is not affected by incomplete foreground subtraction, and can be considered reliable.
The fall-off of the $\dnds$ when $l_{\rm max} > 100$ is expected 
since $l_{\rm max} = 100$
corresponds to an angular scale of the order of $2^\circ$, 
which starts to be compatible with the size of (PSF-smeared) point sources. Therefore, a cleaning with $l_{\rm max} > 100$ starts to remove point sources instead of galactic foregrounds.

\begin{figure}[t]
\centering
\includegraphics[width=1.0\textwidth]{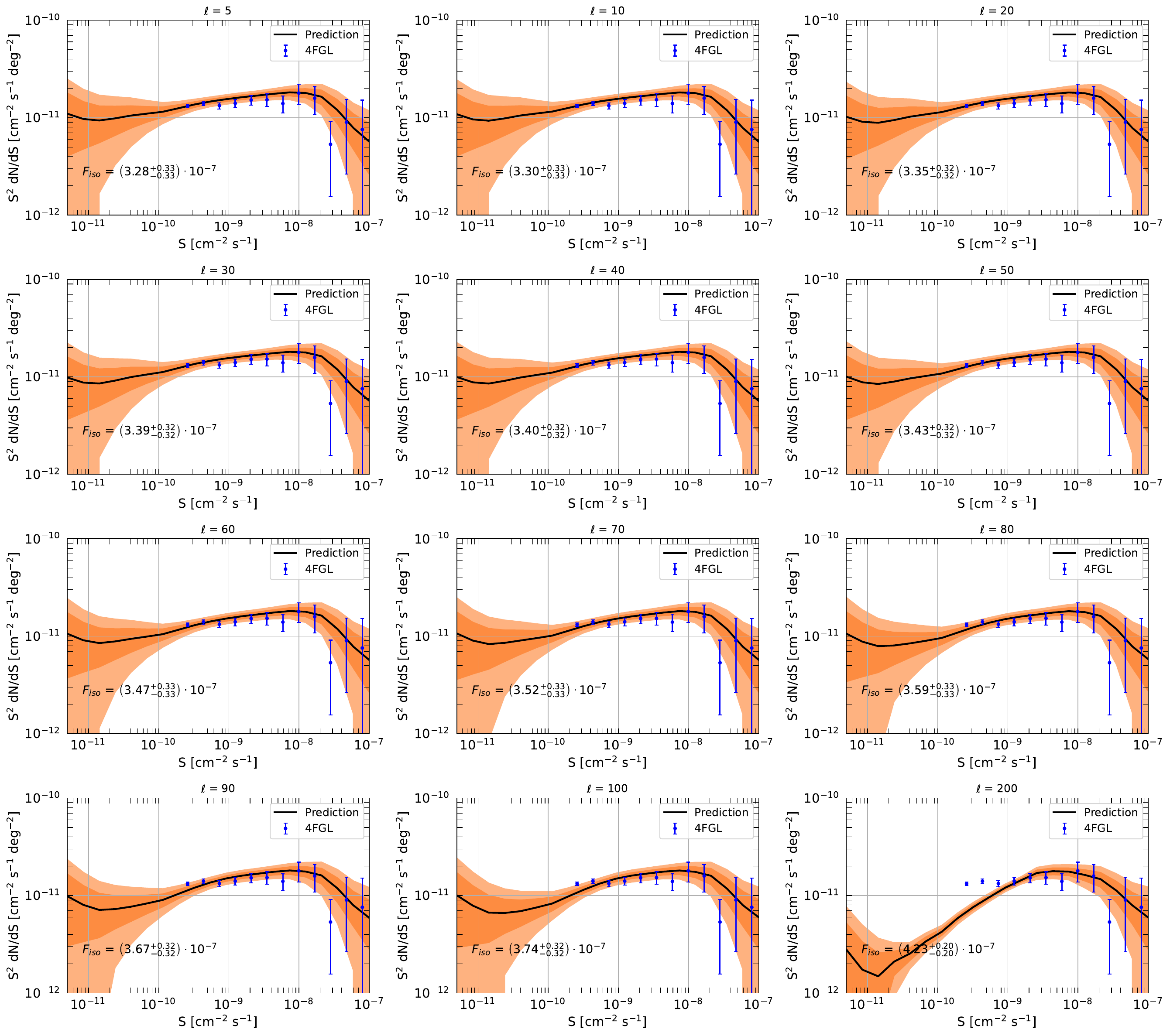}
\caption{Same as Fig. \ref{fig:multipole-cleaningv7} but using a CNN trained with the v05 version of the foreground model, as well as using as input a \fermi count map processed with the same foreground model.}
\label{fig:multipole-cleaningv5}
\end{figure}

The same analysis has been performed on a pipeline fully based on  \verb|gll_iem_v05|: the result is shown in Fig. \ref{fig:multipole-cleaningv5}, to be compared with Fig. \ref{fig:prediction-fermi-12y-fg-v5}. Also in this case, the results are extremely stable up to $l_{\rm max} \sim 100$, and the same above considerations apply.

\begin{figure}[t]
\centering
\includegraphics[width=1.0\textwidth]{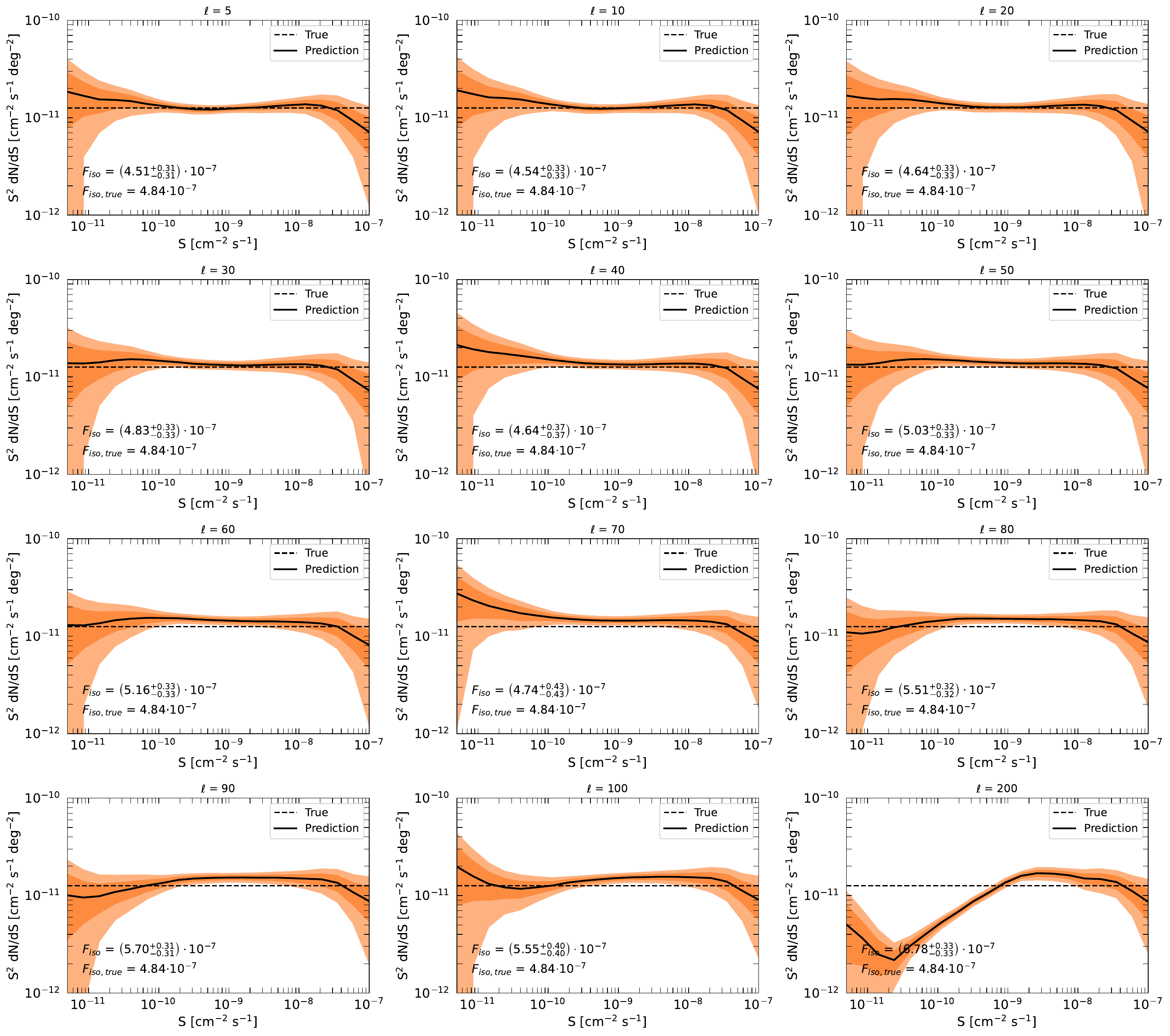}
\caption{Same as Fig. \ref{fig:multipole-cleaningv7} but using as input a synthetic map generated with a flat $S^2 \dnds$ and with the v05 version of the foreground model.}
\label{fig:multipole-cleaning-v5-and-v7}
\end{figure}

Finally, to check the strength of the method against the imperfect knowledge of the background, we performed the same analysis of multipole cleaning by adopting our baseline pipeline (based on \verb|gll_iem_v07|) on maps generated with \verb|gll_iem_v05| (similarly to what has been done for  Fig. \ref{fig:flat-maps-v5}). The results are shown in Fig. \ref{fig:multipole-cleaning-v5-and-v7} , where again we see that the results are pretty consistent, up to $l\sim 100$.
 It can be seen, moreover, that the procedure of multipole cleaning helps in removing the, nonetheless small, positive bias which is present toward low fluxes in this case.  

In conclusion, these tests give confidence on the reliability of the method and that the main result of Fig. \ref{fig:prediction-fermi-12y} corresponds to a fair representation of the source-count distribution of the unresolved gamma-ray sources.

\subsection{Changing $A_{\rm gal}$}
\label{sec:agal-var}

In order to test the stability of the results against changes in the values of $A_{\rm gal}$, we have provided, as input to our trained CNN, \fermi maps cleaned with different values of $A_{\rm gal}$, ranging from 0.78 to 1.0 (our fiducial value is $0.888 \pm 0.005$, thus the chosen range encompass many standard deviations around its central value). The results are shown in Fig. \ref{fig:Agal_change} below. It can be seen that the $dN/dS$ is considerably robust and stable to different values of $A_{\rm gal}$, despite the large range explored. Instead, different values of $A_{\rm gal}$ seem to imply a significant variation in the reconstructed value of $F_{\rm iso}$, as expected since at the large Galactic latitudes explored in our analysis ($|b| > 30^\circ$) there is a certain amount of degeneracy between a purely isotropic emission and the Galactic foreground emission itself. However, this does not affect the reconstructed $dN/dS$.

\subsection{Fully spherical CNN}
\label{sec:full-spherical-nn}

In this section we show a comparison of the results obtained using a fully spherical CNN, instead of the \verb|map2patch| algorithm used in the main analysis. In particular, we adopt here the NNHealpix \cite{nnhealpix} algorithm.

\begin{figure}[t]
\centering
\includegraphics[width=0.9\textwidth]{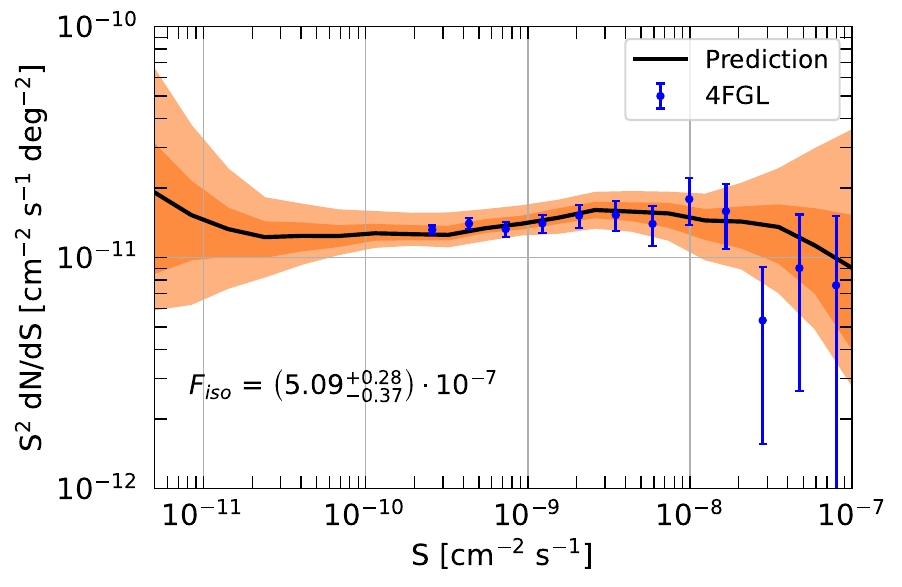}
\caption{Reconstructed $\dnds$ and $\fiso$  and their errors, using a fully spherical  CNN, trained on simulated maps and then applied to the \fermi map. }
\label{fig:SphericalNN-Fermi-dnds}
\end{figure}

Fig. \ref{fig:SphericalNN-Fermi-dnds} shows the reconstructed $\dnds$, $\fiso$  and their frequentist errors, using the spherical CNN applied to the \fermi map. The results are fully compatible with the ones obtained with the \verb|map2patch| algorithm, both for the mean value and the errors, as can be seen by confronting with Fig. \ref{fig:prediction-fermi-12y-bayesian}.

\begin{figure}[t]
\centering
\includegraphics[width=1.0\textwidth]{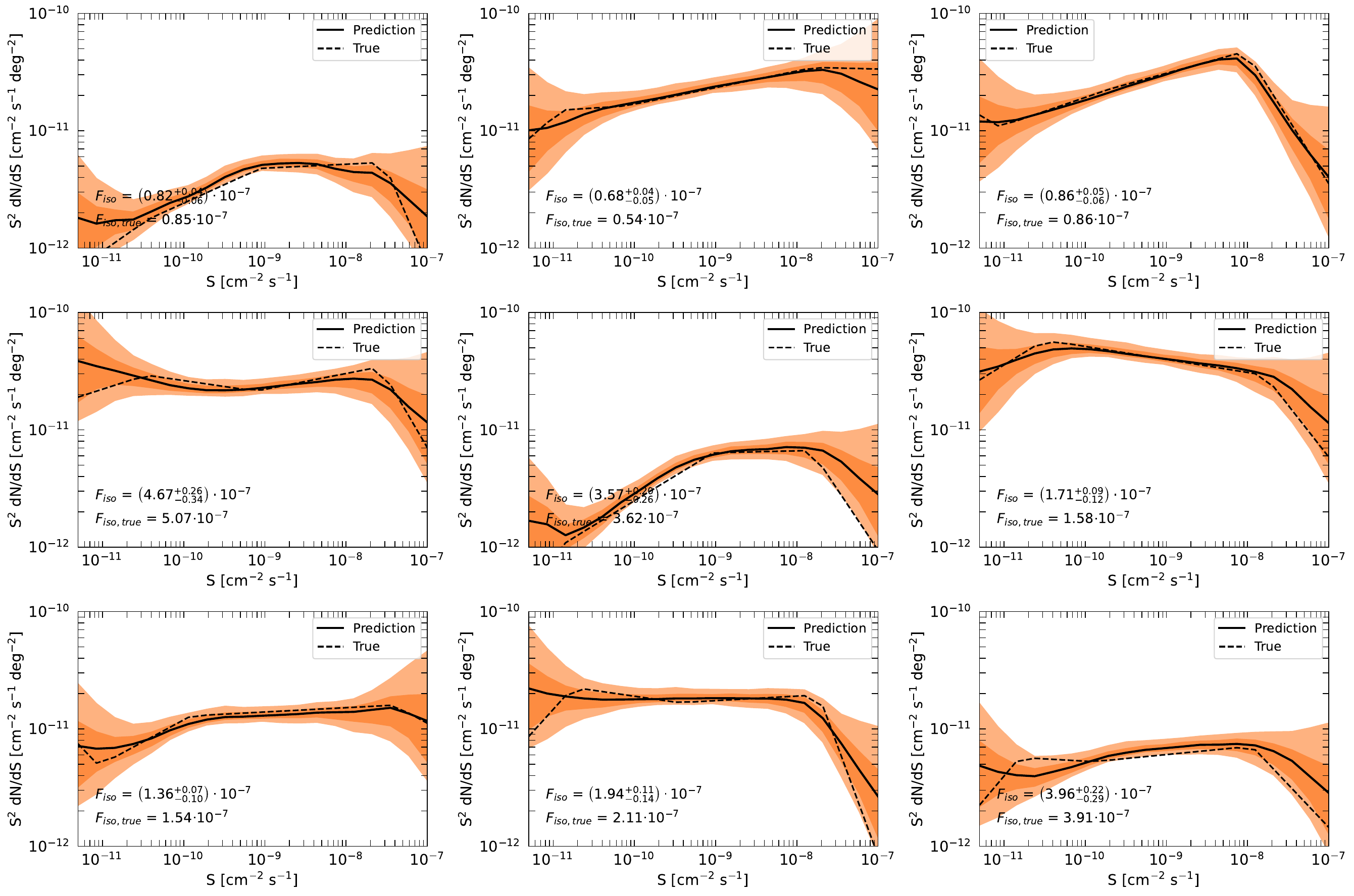}
\caption{Some example $\dnds$ from the validation dataset of the spherical CNN. Each panel shows the input $\dnds$ and $\fiso$ and the corresponding quantities reconstructed by the CNN. The colored areas indicate the $1\sigma$ and $2\sigma$ uncertainty bands. }
\label{fig:SphericalNN-test-random}
\end{figure}

\begin{figure}[t]
\centering
\includegraphics[width=1.0\textwidth]{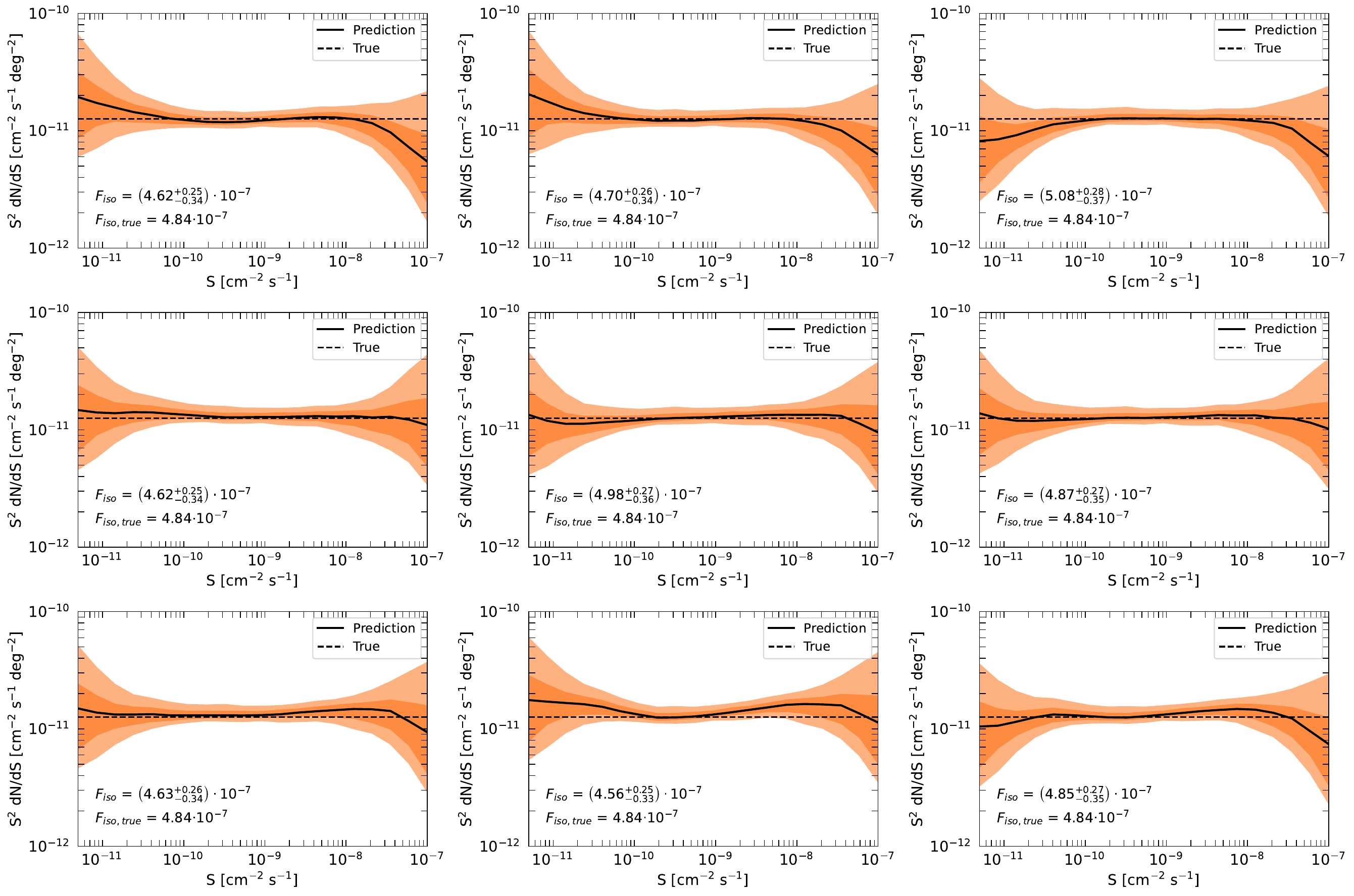}
\caption{Same as Fig. \ref{fig:SphericalNN-test-random}  but for 6 different random realization of a flat $S^2\dnds$ and same 
$F_{\rm iso}$.}
\label{fig:SphericalNN-test-flat}
\end{figure}

Fig. \ref{fig:SphericalNN-test-random} instead, shows some examples of random input $\dnds$ and $\fiso$ from the validation dataset of the Spherical CNN, and the corresponding quantities and their errors reconstructed by the CNN.
The spherical CNN has similar performances as the  CNN that uses \verb|map2patch|, as can be seen comparing with Fig. \ref{fig:random-maps}. The same occurs also in Fig. \ref{fig:SphericalNN-test-flat}, which refers to a flat $S^2 dN/dS$, to be compared with Fig. \ref{fig:flat-maps} obtained with \verb|map2patch|.


\begin{figure}[t]
\centering
\includegraphics[width=0.45\textwidth]{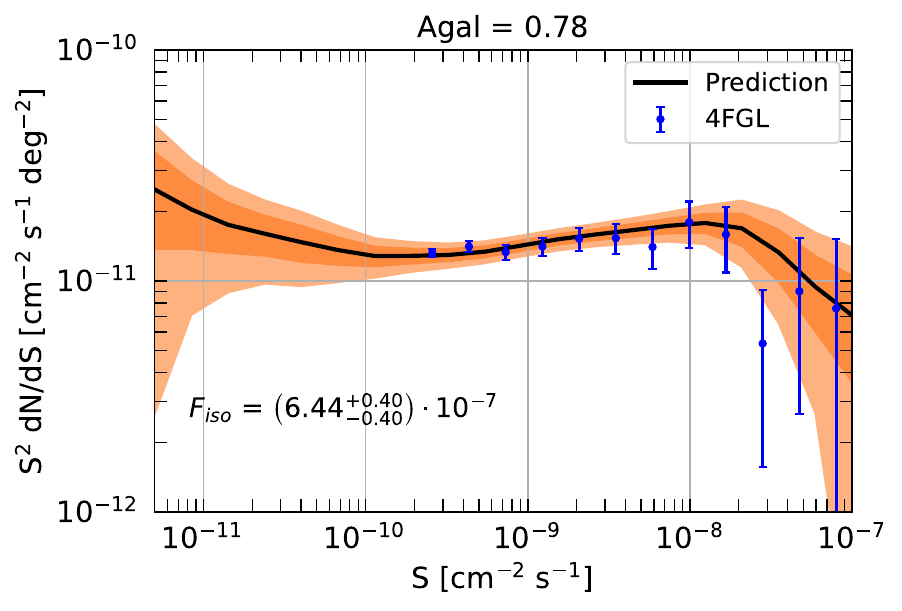}
\includegraphics[width=0.45\textwidth]{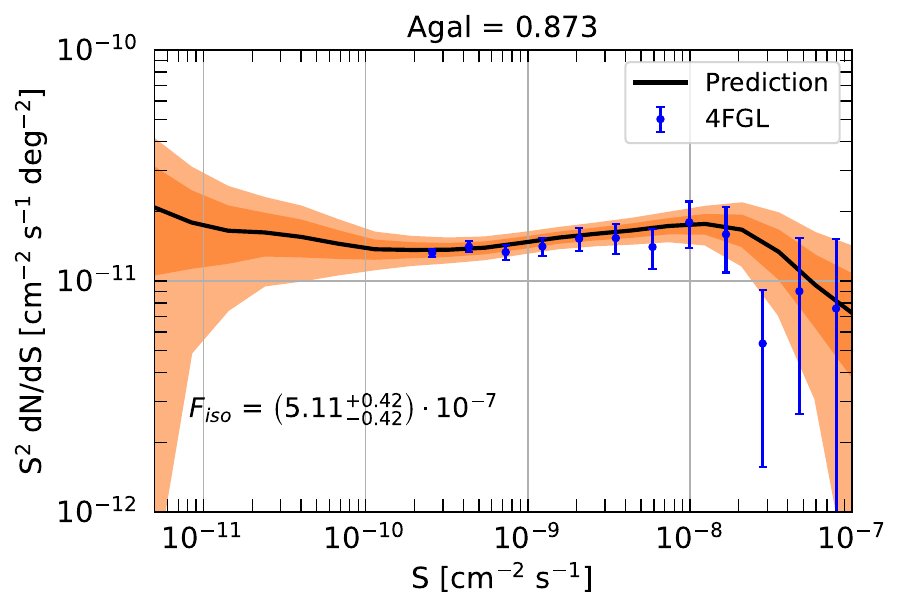}
\includegraphics[width=0.45\textwidth]{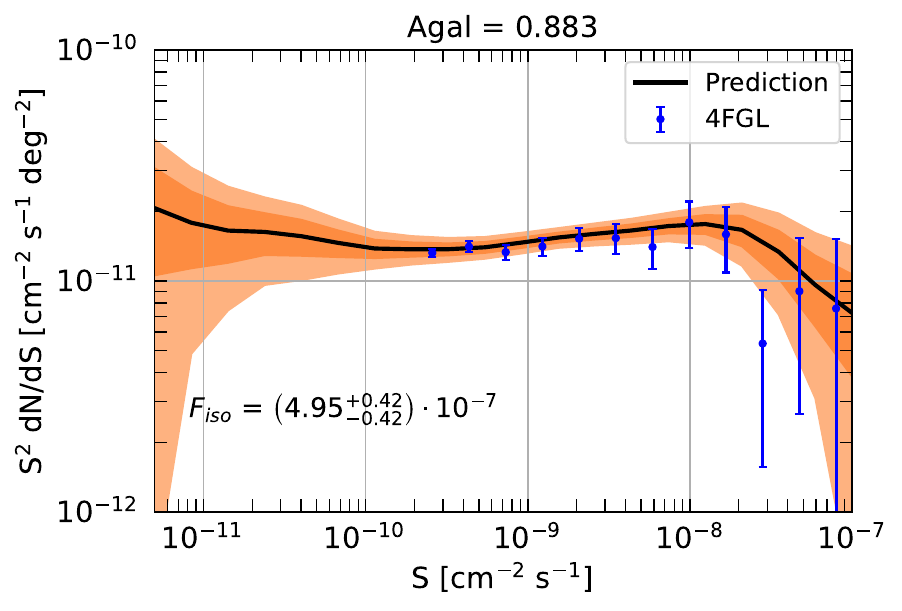}
\includegraphics[width=0.45\textwidth]{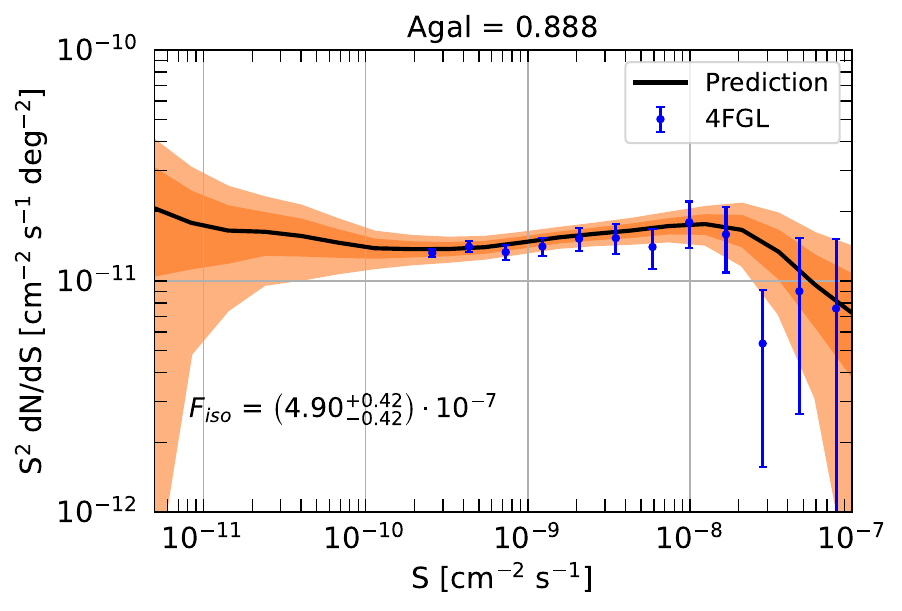}
\includegraphics[width=0.45\textwidth]{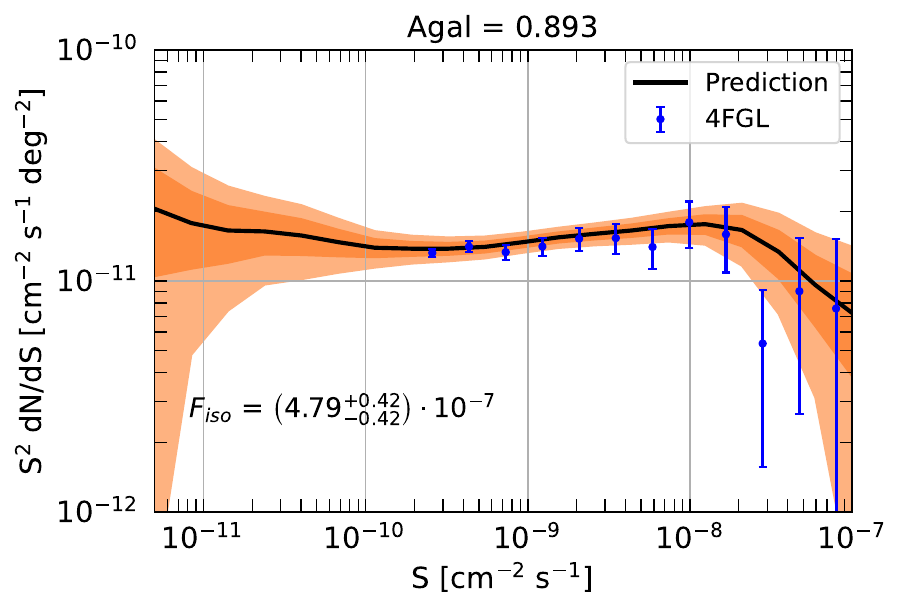}
\includegraphics[width=0.45\textwidth]{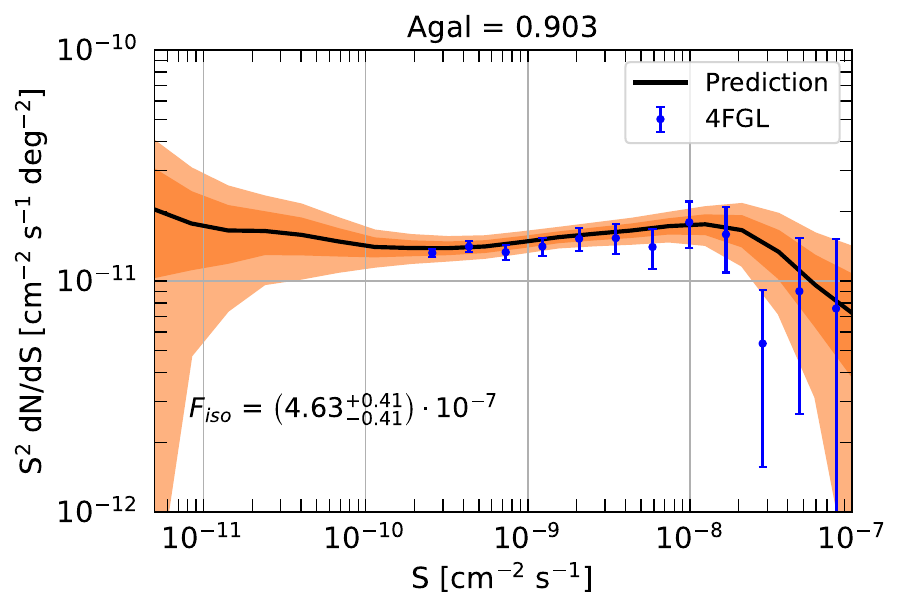}
\includegraphics[width=0.45\textwidth]{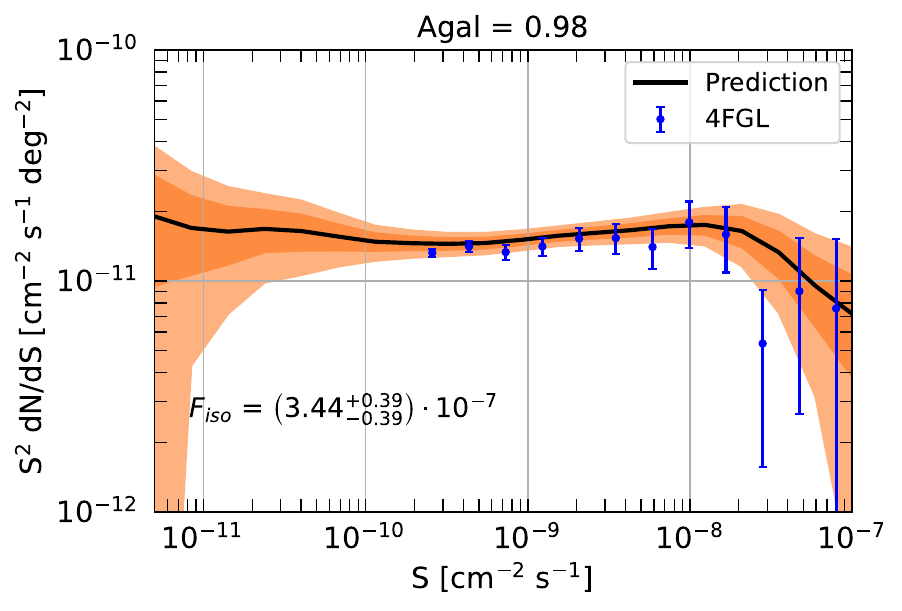}
\includegraphics[width=0.45\textwidth]{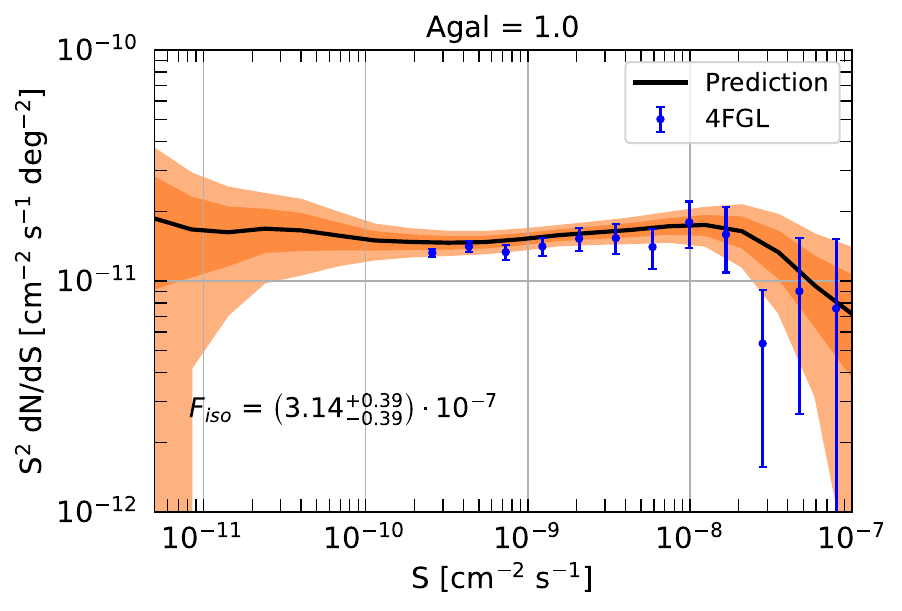}
\caption{
Reconstructed $\dnds$ and $\fiso$  and their Bayesian errors, when the trained CNN is applied to the \fermi map  using  different values of $A_{\rm gal}$ as indicated in the panel labels.}
\label{fig:Agal_change}
\end{figure}

\end{document}